\documentclass[prb,superscriptaddress,showpacs,floatfix,tightenlines,10pt,twocolumn]{revtex4-1}
\usepackage{amssymb}
\usepackage{amsmath}
\usepackage{graphicx}
\usepackage{float}
\usepackage{xcolor}

\newcommand{\func}[1]{\operatorname{#1}}

\begin{document}

\title{Chiral filtration and Faraday rotation in multi-Weyl semimetals}
\author{Ren\'{e} C\^{o}t\'{e}}
\author{R\'{e}mi N. Duchesne}
\author{Gautier D. Duchesne}
\author{Olivier Tr\'{e}panier}
\affiliation{D\'{e}partement de physique and Institut Quantique, Universit\'{e} de
Sherbrooke, Sherbrooke, Qu\'{e}bec, Canada J1K 2R1 }
\date{\today }

\begin{abstract}
In Weyl semimetals with broken inversion and time-reversal symmetries, the
Maxwell equations are modified by the presence of the axion terms $\mathbf{b}
$ and $b_{0}$ where, in the simplest case of a two-node Weyl semimetal, $%
2\hslash \mathbf{b}$ is the vector that connects two Weyl nodes in momentum
space and $2\hslash b_{0}$ is the separation in energy of the two Dirac
points of these nodes. These axion terms modify the behavior of
electromagnetic waves inside a Weyl semimetal leading to a number of unique
optical properties such as non-reciprocal propagation, circular and linear
dichroism, birefringence and Faraday and Kerr rotations in the absence of a
magnetic field. These effects can be used to design optical devices that act
as broadband chiral filters, circular polarizers or tunable optical
isolators. In this paper, we study in detail how the Faraday and Kerr
rotations as well as the transmission and reflection of light incident on a
slab of Weyl semimetal can be controlled by varying the different parameters
characterizing the Weyl semimetal such as the axion terms, the Fermi level
and Fermi velocity, the background dielectric constant, the scattering time
for intraband scattering, the width of the semimetal and the dielectric
constant of the dielectrics on each side of the semimetal slab. We extend
our analysis to Weyl nodes with Chern number $n=1,2,3$.
\end{abstract}

\maketitle

\section{INTRODUCTION}

According to the Nielsen-Ninomiya theorem\cite{Nielsen}, the simplest model
of a Weyl semimetal with broken inversion and time-reversal symmetries
consists of two Weyl nodes of opposite topological charge (or Chern number), 
$n=\pm 1,$ separated in momentum space by $2\hslash \mathbf{b}$ and in
energy by $2\hslash b_{0}$. Near these nodes, the electronic dispersion is
linear in wave vector in all three directions of momentum space \textit{i.e.}
$E_{s}=s\hslash v_{F}\left\vert \mathbf{k}\right\vert ,$ where the wave
vector $\mathbf{k}$ is measured with respect to the Weyl point and $s=\pm 1$
is the band index (the upper and lower part of the Dirac cone). Density
functional calculations suggest that more complex WSMs could host Weyl nodes
with higher topological charge $n.$ In these so-called multi-Weyl semimetals
(mWSMs), the dispersion remains linear along the direction parallel to the
vector $\mathbf{b}$ but becomes quadratic (for $n=2$) or cubic (for $n=3$)
in the perpendicular direction. The maximum topological charge, $n=3,$ is
dictated by the discrete rotational symmetry on the lattice\cite%
{FanBernevig2012}.

The topological nature of the band structure of WSMs leads to unusual
transport properties such as an anomalous Hall effect\cite{AHE}, a
chiral-magnetic effect\cite{CME}, Fermi arcs on the surfaces parallel to the
vector $\mathbf{b}$\cite{FermiArc}, a chiral anomaly leading to a negative
longitudinal negative magnetoresistance\cite{ChiralAnomaly}, a plasmon\cite%
{Spivak} whose frequency increases as $\sqrt{B}$ in a magnetic field $B$.
These properties have been the subject of numerous reviews\cite{Review}.

Weyl semimetals also exhibit a number of surprising optical properties
because the Maxwell equations that govern the propagation of electromagnetic
waves in a WSM are modified by the inclusion of the axion terms $\mathbf{b}$
and $b_{0}$\cite{Wilczek,Wu, Burkov2012}. These terms generate an extra
current $\mathbf{j}=-\frac{e^{2}}{2\pi ^{2}\hslash }\mathbf{b}\times \mathbf{%
E}+\frac{e^{2}}{2\pi ^{2}\hslash }b_{0}\mathbf{B}$ in the Amp\`{e}re-Maxwell
equation and an extra density $\rho =\frac{2\alpha }{\pi }\sqrt{\frac{%
\varepsilon _{0}}{\mu _{0}}}\mathbf{b}\cdot \mathbf{B}$ in the Gauss law ($%
\alpha =e^{2}/4\pi \varepsilon _{0}\hslash c$ is the fine-structure constant
and $\varepsilon _{0},\mu _{0}$ are the permittivity and permeability of
free space). The modified Maxwell equations lead, in the absence of an
external magnetic field, to giant optical nonreciprocity, circular and
linear birefringence and dichroism and Faraday and Kerr rotations\cite%
{Kargarian,Cote2022}. Conversely, a study of the electromagnetic eigenmodes
can give information about the band structure of a WSM such as the energy
and momentum separations of the Weyl nodes and the position of the Fermi
level\cite{ChenEM2019}.

In the Faraday configuration where the light propagates in the direction of
the axion term $\mathbf{b}$ and in the absence of a magnetic field, the
solutions of the Maxwell equations are forward and backward right-circularly
(RCP)\ and left-circularly polarized (LCP) waves. When the propagation is
perpendicular to $\mathbf{b}$ (the Voigt configuration), the solutions are
linearly polarized waves. It has previously been shown\cite{Berman2021} that
a WSM can serve as a broadband chiral optical medium that selectively
transmits and reflects circularly polarized electromagnetic wave in a wide
range of frequencies when the wave vector $\mathbf{q}$ of the
electromagnetic wave is parallel to the axion term $\mathbf{b}$. When $%
\mathbf{q}$ is perpendicular to $\mathbf{b},$ the WSM\ acts instead as a
linear polarization filter, transmitting light when the polarization vector
is collinear with $\mathbf{b}$ and reflecting it otherwise. This chiral
filtration effect has been considered in some recent papers where it was
shown that the frequency range where the right or left circularly polarized
light propagates can be controlled, to some extend, by changing the value of
the axion term $\mathbf{b}$ and/or the doping of the Weyl semimetal allowing
the WSM\ to be used as a broadband polarizer and making the absorption
perfectly tunable\cite{Wu2023,Yang2022,Ghosh2023}. These special optical
properties of WSMs may be used in photonic applications and devices\cite%
{Yang2022,Guo2023, Asadchy2020}.

In this paper, we study in detail how the transmission, reflection and Kerr
and Faraday rotations of an electromagnetic wave incident on a slab of a
Weyl semimetal are modified when the different parameters characterizing the
WSM are varied. Our goal is to better understand the tunability of the
chiral filtering and polarizing effects of WSMs. The list of parameters that
we consider includes the polarization and frequency of the incident wave,
the value of the axion terms $\mathbf{b}$ and $b_{0},$ the thickness $d$ and
Fermi level $e_{F}$ (doping) of the WSM, the Fermi velocity $v_{F}$, the
relaxation time $\tau $ for intraband transitions and the value of the
background dielectric constant $\varepsilon _{b}$ and cutoff wave vector $%
k_{c}.$ We also consider the situation where the WSM is sandwiched between
two dielectric media with different refractive indices $n_{1}$ and $n_{3}$
and extent our analysis to Weyl nodes with higher Chern numbers $n=2,3$. 
\textbf{\ }Since we work with a slab of WSM, we discuss the transmission
resonances that occur when the wavelength of light in the WSM\ is comparable
with the thickness $d$ so that the WSM\ acts as a Fabry-P\'{e}rot
interferometer. We extend our analysis to the thin-film limit $d<<\lambda ,$
where $\lambda $ is the wavelength of light in the WSM and to the limit $%
d\rightarrow \infty $ where the WSM is semi-infinite. Our study complements
recent works on the subject\cite{Gupta2022,Ghosh2023} by giving a more
in-depth study of the effects of the different parameters characterizing the
Weyl semimetal on some of its optical properties.

The remainder of this paper is organized as follows. In Sec. II, we define a
simple model for the two-node WSM and compute the corresponding dielectric
tensor. Section III\ is devoted to the derivation of the dispersion relation
and polarization of the propagating waves inside the WSM when the
propagation is along the vector connecting both nodes. In Sec. IV, we apply
the Maxwell boundary conditions to derive the transmission and reflection
coefficients for the RCP, LCP\ and linearly polarized (LP) waves and the
Faraday and Kerr rotation and ellipticity angles of the transmitted and
reflected waves. Our numerical results are presented and discussed in Sec.
V. We conclude in Sec. VI.

\section{HAMILTONIAN OF A MULTI-WEYL SEMIMETAL}

We consider a two-node model of a Weyl semimetal with broken inversion and
time-reversal symmetries so that the two nodes of opposite chiralities are
separated in momentum space and in energy. The nodes are centered at wave
vectors $\chi b\widehat{\mathbf{z}}$ in the Brillouin zone and separated in
energy by $2\hslash b_{0}$, where $\chi =\pm 1$ stands for the node index.
We assume that the two nodes are untilted and that there is no external
magnetic field. In the continuum approximation, which is valid for small
wave vector deviations from a Weyl node, the Hamiltonian of an electron in
node $\chi $ with Chern number $C=n,$ with $n=1,2,3,$ is given by\cite%
{FanBernevig2012} 
\begin{equation}
h_{\chi }\left( \mathbf{k}\right) =\chi \hslash v_{F}\left[ k_{z}\sigma
_{z}+\beta \left( k_{-}^{n}\sigma _{+}+k_{+}^{n}\sigma _{-}\right) \right]
+\chi \hslash b_{0}I_{2},  \label{hamilton}
\end{equation}%
where $v_{F}$ is the Fermi velocity for the dispersion in $k_{z}$, $\beta $
characterizes the anisotropy in the dispersion ($\beta $ has units m$^{n-1}$%
), $\mathbf{k}$ is the wave vector measured from the position of the Weyl
node, $\mathbf{\sigma }$ is the vector of Pauli matrices defined in the
basis of the two electronic bands that cross and $I_{2}$ is the $2\times 2$
unit matrix. We have also defined $\sigma _{\pm }=\sigma _{x}\pm i\sigma
_{y} $ and $k_{\pm }=k_{x}\pm ik_{y}.$ The terms $\mathbf{b}$ and $b_{0}$
break time-reversal and space-inversion symmetry respectively. According to
Ref. \onlinecite{FanBernevig2012}: "$C_{4}$ or $C_{6}$ symmetry in a crystal
can protect double-Weyl nodes while only $C_{6}$ symmetry can protect
triple-Weyl nodes. There cannot be any higher order crossings protected by $%
n-$fold rotation symmetries".

The electronic energy spectrum for each node is given in cylindrical
coordinates by%
\begin{equation}
E_{s}\left( \mathbf{k}\right) =s\hslash v_{F}\sqrt{k_{z}^{2}+\beta
^{2}k_{\bot }^{2n}}+\chi \hslash b_{0},
\end{equation}%
where $s=\pm 1$ is the band index. The corresponding eigenstates are given by%
\begin{equation}
\left\vert \eta _{s,\chi }\left( \mathbf{k}\right) \right\rangle =\frac{1}{%
h_{s,\chi }}\left( 
\begin{array}{c}
k_{z}+\chi s\sqrt{k_{z}^{2}+\beta ^{2}k_{\bot }^{2n}} \\ 
\beta k_{\bot }^{n}e^{in\varphi }%
\end{array}%
\right) ,  \label{eigenv}
\end{equation}%
where $h_{s,\chi }$ is a normalization constant. We assume that the two
nodes are at a thermodynamical equilibrium at $T=0$ K and share a common
Fermi level $e_{F}$ so that the local Fermi level energy $e_{F,\chi }$ in
each node is given by 
\begin{equation}
e_{F,\tau }=e_{F}-\chi \hslash b_{0}.
\end{equation}

For Chern number $n=1,$the Hamiltonian is isotropic in $\mathbf{k}$ around
each Weyl node so that the (relative) dielectric tensor is isotropic, $%
\varepsilon _{ij}\left( \omega \right) =\varepsilon \left( \omega \right)
\delta _{ij}.$ This is no longer true if $n\neq 1$ and we have instead\cite%
{Conductivite} that $\varepsilon _{xx}=\varepsilon _{yy}=\varepsilon \left(
\omega \right) $ with $\varepsilon \left( \omega \right) $ given by%
\begin{eqnarray}
\varepsilon \left( \omega \right)  &=&\varepsilon _{b}  \label{epsilon} \\
&&+\gamma \sum_{\chi }\left[ i\theta \left( \omega -\frac{2e_{F,\chi }}{%
\hslash }\right) +\frac{1}{\pi }\ln \left( \left\vert \frac{\omega
^{2}-4v_{F}^{2}k_{c}^{2}}{\omega ^{2}-4e_{F,\chi }^{2}/\hslash ^{2}}%
\right\vert \right) \right]   \notag \\
&&+\gamma \sum_{\chi }\frac{4}{\pi \hslash ^{2}\omega }\frac{i\tau }{%
1+\omega ^{2}\tau ^{2}}\left( e_{F,\chi }^{2}+\frac{1}{3}\hslash ^{2}\omega
^{2}+\frac{\hslash ^{2}}{4\tau ^{2}}\right)   \notag \\
&&-\gamma \sum_{\chi }\frac{4}{\pi \hslash ^{2}}\frac{\tau ^{2}}{1+\omega
^{2}\tau ^{2}}\left( e_{F,\chi }^{2}-\frac{\hslash ^{2}}{12\tau ^{2}}\right)
,  \notag
\end{eqnarray}%
where we have defined%
\begin{equation}
\gamma =\frac{n\alpha }{6v_{F}/c}
\end{equation}%
and  $\alpha =e^{2}/4\pi \varepsilon _{0}\hslash c$ is the fine-structure
constant with $\varepsilon _{0},\mu _{0}$ the permittivity and permeability
of free space, $k_{c}$ is a higher-energy cutoff wave vector, and $%
\varepsilon _{b}$ is the (relative) background dielectric constant that can
be large in WSMs. For example $\varepsilon _{b}=6.2$ in WSM TaAs\cite%
{Kotov2016}. The dependence of $\varepsilon _{zz}\left( \omega \right) $ on
the Chern number $n$ is more complex\cite{Gupta2022} and we do not give it
here since it is not needed in any of the calculations in this paper.

The first (last) two lines in Eq. (\ref{epsilon}) are contributions from the
interband (intraband) transitions with $\tau $ the relaxation time for the
intraband transitions. Both contributions, for Chern number $n=1,$ have been
calculated in parts in a number of papers\cite%
{Kargarian,Sonowal2019,Carbotte2014,Carbotte2016,Carbotte2018,Carbotte2021,Cote2022}%
. It is easy to show\cite{Conductivite} that the only difference for Chern
number $n$ is a multiplication of the conductivities $\sigma _{xx}$ and $%
\sigma _{yy},$ calculated with $n=1,$ by the Chern number $n.$ For
completeness, we give a proof of this statement in Appendix A. The
off-diagonal conductivity $\sigma _{xy}\left( \omega \right) $ scales with $%
n $ in the same way. To order $\omega ^{2}$, it is given by\cite%
{Ghosh2023,Conductivite} 
\begin{equation}
\func{Re}\left[ \sigma _{xy}\left( \omega \right) \right] =\frac{ne^{2}b}{%
\pi h}\left( 1+\frac{1}{12}\frac{\omega ^{2}}{%
v_{F}^{2}k_{c}^{2}-v_{F}^{2}b^{2}}\right) .  \label{axion}
\end{equation}%
With the parameters that we use in this paper, the $\omega ^{2}$ correction
is negligible and we ignore it in our calculations.

The density of states measured with respect to the Dirac point in each node
is defined by $g\left( E\right) =\frac{1}{V}\sum_{\mathbf{k}}\delta \left(
E-E\left( \mathbf{k}\right) \right) $. With $E\left( \mathbf{k}\right)
=\hslash v_{F}\sqrt{k_{z}^{2}+\beta ^{2}k_{\bot }^{2n}},$ it is given for $%
n=1,2,3$ by $g_{n}\left( E\right) $ where 
\begin{eqnarray}
g_{1}\left( E\right) &=&\frac{E^{2}}{2\pi ^{2}\beta ^{2}\hslash ^{3}v_{F}^{3}%
}, \\
g_{2}\left( E\right) &=&\frac{E}{8\pi \beta \hslash ^{2}v_{F}^{2}}, \\
g_{3}\left( E\right) &=&\frac{E^{2/3}}{12\beta ^{2/3}\pi ^{3/2}\left(
\hslash v_{F}\right) ^{5/3}}\frac{\Gamma \left( \frac{1}{3}\right) }{\Gamma
\left( \frac{5}{6}\right) }.
\end{eqnarray}%
The total electronic density $n_{e,n}$ in the WSM\ is thus related to the
Fermi level $e_{F}$ by%
\begin{eqnarray}
n_{e,1} &=&e_{F}\frac{3\hslash ^{2}b_{0}^{2}+e_{F}^{2}}{3\pi ^{2}\beta
^{2}\hslash ^{3}v_{F}^{3}}, \\
n_{e,2} &=&\frac{\hslash ^{2}b_{0}^{2}+e_{F}^{2}}{8\pi \beta \hslash
^{2}v_{F}^{2}},\text{ } \\
n_{e,3} &=&\frac{\left( \left( e_{F}+\hslash b_{0}\right) ^{5/3}+\left(
e_{F}-\hslash b_{0}\right) ^{5/3}\right) \Gamma \left( \frac{1}{3}\right) }{%
20\beta ^{2/3}\pi ^{3/2}\left( \hslash v_{F}\right) ^{5/3}\Gamma \left( 
\frac{5}{6}\right) }.
\end{eqnarray}

\section{ELECTROMAGNETIC MODES}

The Maxwell equations in a WSM with Chern number $n=1$\ are modified by the
presence of the axion field $\theta \left( \mathbf{r},t\right) =2\mathbf{b}%
\cdot \mathbf{r}-2b_{0}t$ and become\cite{Burkov2012}%
\begin{eqnarray}
\nabla \cdot \mathbf{D} &=&\rho _{f}+\frac{2\alpha }{\pi }\sqrt{\frac{%
\varepsilon _{0}}{\mu _{0}}}\mathbf{b}\cdot \mathbf{B,} \\
\nabla \cdot \mathbf{B} &=&0, \\
\nabla \times \mathbf{E} &=&-\frac{\partial \mathbf{B}}{\partial t}, \\
\nabla \times \mathbf{H} &=&\frac{\partial \mathbf{D}}{\partial t}+\mathbf{J}%
_{f}-\frac{2\alpha }{\pi }\sqrt{\frac{\varepsilon _{0}}{\mu _{0}}}\left( 
\mathbf{b}\times \mathbf{E}-b_{0}\mathbf{B}\right) \mathbf{.}
\end{eqnarray}%
In these equations, $\rho _{f}$ is the free (or induced) charge density and $%
\mathbf{J}_{f}=\sigma \left( \omega \right) \mathbf{E}$ is the induced
current density with the conductivity obtained from $\varepsilon \left(
\omega \right) =\varepsilon _{b}+i\sigma \left( \omega \right) /\varepsilon
_{0}\omega $ with $\varepsilon \left( \omega \right) $ given in Eq. (\ref%
{epsilon}). We assume that the relative permeability $\mu _{r}=\mu /\mu _{0}$
is unity. In mWSMs, Eq. (\ref{axion}) shows that $\mathbf{b}$ is replaced by 
$n\mathbf{b}$ in the Maxwell equations. This also true for $b_{0}$ which is
replaced by $nb_{0}$\cite{Termenb0}. The gyrotropic effect of $b_{0}$ is
very small in comparison with that due to $\mathbf{b}$ so that we neglect
its presence in the Maxwell equations\cite{Cote2022}. We keep $b_{0}$ in the
calculation of $\sigma \left( \omega \right) ,$ however, since it determines
the density of electrons in each node.

The wave equation can be written as $M_{i,j}E_{j}=0$ where the matrix $M$ is
defined by

\begin{equation}
M_{ij}\left( \omega \right) =-c^{2}\left( q^{2}\delta
_{ij}-q_{i}q_{j}\right) +\omega ^{2}\widetilde{\varepsilon }_{ij},
\label{matricem}
\end{equation}%
where $\delta _{ij}$ is the Kronecker delta and the effective dielectric
tensor $\widetilde{\varepsilon }_{ij}$ is given by%
\begin{equation}
\widetilde{\varepsilon }_{ij}\left( \omega \right) =\varepsilon \left(
\omega \right) \delta _{i,j}+\frac{2in\alpha c}{\pi \omega }\varepsilon
_{ijk}b_{k},  \label{epsitilde}
\end{equation}%
where $\varepsilon _{ijk}$ is the Levi-Civita symbol. The axion term $%
\mathbf{b}$ thus appears as an off-diagonal term in the effective dielectric
tensor of a WSM.

In this paper, we work in the so-called Faraday configuration: we take $%
\mathbf{b}=b\widehat{\mathbf{z}}$ and consider a wave propagating along the $%
z$ axis so that we have, for the matrix $M,$ the simple form%
\begin{equation}
M=\left( 
\begin{array}{ccc}
\omega ^{2}\varepsilon \left( \omega \right) -c^{2}q^{2} & in\omega \kappa cb
& 0 \\ 
-in\omega \kappa cb & \omega ^{2}\varepsilon \left( \omega \right)
-c^{2}q^{2} & 0 \\ 
0 & 0 & \omega ^{2}\varepsilon _{zz}\left( \omega \right) 
\end{array}%
\right) ,  \label{simple}
\end{equation}%
with the newly defined constant $\kappa =2\alpha /\pi .$ The dispersion
relations of the electromagnetic modes are given by 
\begin{eqnarray}
q_{1,\pm } &=&\pm \frac{\omega }{c}\sqrt{\varepsilon _{+}\left( \omega
\right) },  \label{q1} \\
q_{2,\pm } &=&\pm \frac{\omega }{c}\sqrt{\varepsilon _{-}\left( \omega
\right) },  \label{q2}
\end{eqnarray}%
where the dielectric functions%
\begin{equation}
\varepsilon _{\pm }\left( \omega \right) =\varepsilon \left( \omega \right)
\pm \frac{nc\kappa b}{\omega }  \label{epsipm}
\end{equation}%
give the complex refractive indices%
\begin{equation}
n_{\pm }\left( \omega \right) =\sqrt{\varepsilon _{\pm }\left( \omega
\right) }.
\end{equation}

The polarization vectors $\mathbf{e}_{+}$ for the modes $q_{1,\pm }$ and $%
\mathbf{e}_{-}$ for the modes $q_{2,\pm }$ are given by 
\begin{equation}
\mathbf{e}_{+}=\frac{1}{\sqrt{2}}\left( 
\begin{array}{c}
i \\ 
1 \\ 
0%
\end{array}%
\right) ;\,\mathbf{e}_{-}=\frac{1}{\sqrt{2}}\left( 
\begin{array}{c}
-i \\ 
1 \\ 
0%
\end{array}%
\right) .  \label{vec1}
\end{equation}

The wave vectors $q_{1,+}\left( q_{1,-}\right) $ correspond to the
forward(backward) propagation of a right(left)-circularly polarized (RCP and
LCP) wave while $q_{2,+}\left( q_{2,-}\right) $ correspond to the forward
(backward) propagation of a left (right)-circularly polarized wave. For a
given circular polarization, the dielectric function is different for
forward and backward propagations \textit{i.e. }the axion term leads to
non-reciprocal propagation of the electromagnetic waves. Moreover, the
dielectric function is different for RCP and LCP\ wave so that the WSM
exhibits circular birefringence as well as circular dichroism since the two
waves are attenuated differently. Dispersion relations for other direction
of propagation have also been derived\cite{ChenEM2019}.

\section{TRANSMISSION AND REFLECTION COEFFICIENTS}

In order to calculate the transmission and reflection coefficients, we
consider a monochromatic electromagnetic wave with amplitude $E_{0}$
impinging at normal incidence on the surface of a WSM that we take at $z=0$
and propagating along the direction of the axion vector $\mathbf{b}=b%
\widehat{\mathbf{z}}\mathbf{.}$ This Faraday configuration is the simplest
to study since there is no Fermi arc on the surfaces perpendicular to the
vector $\mathbf{b}.$ The slab of WSM\ occupies region 2 defined by $z\in %
\left[ 0,d\right] ,$ where $d$ is the width of the WSM. Medium $1$ ($z<0$)
and $3$ ($z>d$) are dielectrics with refractive indices $n_{1}$ and $n_{3}$
respectively. The electric and magnetic fields in regions $1$ and $3$ are
given by

\begin{eqnarray}
\mathbf{E}_{1}\left( z\mathbf{,}t\right) &=&E_{0}\left[ \mathbf{e}%
_{i}e^{i\left( q_{i}z-\omega t\right) }+r\mathbf{e}_{r}e^{-i\left(
q_{i}z+\omega t\right) }\right] , \\
\mathbf{B}_{1}\left( z\mathbf{,}t\right) &=&\frac{q_{i}}{\omega }E_{0}\left[
\left( \widehat{\mathbf{z}}\times \mathbf{e}_{i}\right) e^{i\left(
q_{i}z-\omega t\right) }-r\left( \widehat{\mathbf{z}}\times \mathbf{e}%
_{r}\right) e^{-i\left( q_{i}z+\omega t\right) }\right] ,  \notag
\end{eqnarray}%
and 
\begin{eqnarray}
\mathbf{E}_{3}\left( z\mathbf{,}t\right) &=&E_{0}t\mathbf{e}_{t}e^{i\left(
q_{t}z-\omega t\right) }, \\
\mathbf{B}_{3}\left( z\mathbf{,}t\right) &=&\frac{q_{t}}{\omega }%
E_{0}t\left( \widehat{\mathbf{z}}\times \mathbf{e}_{t}\right) e^{i\left(
q_{t}z-\omega t\right) },  \notag
\end{eqnarray}%
where $r$ and $t$ are reflection and transmission factors and the $\mathbf{e}%
^{\prime }s$ are polarization vectors. The incident and transmitted wave
vectors are given by%
\begin{eqnarray}
q_{i} &=&q_{0}n_{1}, \\
q_{t} &=&q_{0}n_{3},
\end{eqnarray}%
where $q_{0}=\omega /c.$ We write the polarization vectors as%
\begin{eqnarray}
\mathbf{e}_{i} &=&\alpha _{i}\widehat{\mathbf{x}}+\beta _{i}\widehat{\mathbf{%
y}}, \\
\mathbf{e}_{t} &=&\alpha _{t}\widehat{\mathbf{x}}+\beta _{t}\widehat{\mathbf{%
y}}.
\end{eqnarray}

Inside the WSM, there are four propagating waves so that the electric and
magnetic fields are given by%
\begin{eqnarray}
\mathbf{E}_{2}\left( z\mathbf{,}t\right) &=&\sum_{n=1}^{n=4}E_{0}t_{n}%
\mathbf{e}_{n}e^{i\left( q_{n}z-\omega t\right) }, \\
\mathbf{B}_{2}\left( z\mathbf{,}t\right) &=&\sum_{n=1}^{n=4}\frac{q_{1}}{%
\omega }E_{0}t_{n}\left( \widehat{\mathbf{z}}\times \mathbf{e}_{n}\right)
e^{i\left( q_{n}z-\omega t\right) },
\end{eqnarray}%
with the polarization vectors written as%
\begin{equation}
\mathbf{e}_{n}=\alpha _{n}\widehat{\mathbf{x}}+\beta _{n}\widehat{\mathbf{y}}%
.
\end{equation}

The factors $t_{n}$ correspond to forward (F) and backward (B) propagations
along the $z$ axis with RCP\ and LCP polarization as described in Table 1.

\begin{center}
%TCIMACRO{\TeXButton{B}{\begin{table}[tbp] \centering}}%
%BeginExpansion
\begin{table}[tbp] \centering%
%EndExpansion
\begin{tabular}{|l|l|l|}
\hline
$t_{1}$ & $q_{1}=q_{1+},\,\mathbf{e}_{1}=\mathbf{e}_{+}$ & F-RCP \\ \hline
$t_{2}$ & $q_{2}=q_{1-},\,\mathbf{e}_{2}=\mathbf{e}_{+}$ & B-LCP \\ 
$t_{3}$ & $q_{3}=q_{2+},\,\mathbf{e}_{3}=\mathbf{e}_{-}$ & $\text{F-LCP}$ \\ 
$t_{4}$ & $q_{4}=q_{2-},\,\mathbf{e}_{4}=\mathbf{e}_{-}$ & B-RCP \\ \hline
\end{tabular}%
\caption{Description of the coefficients $t_n$.
}\label{tableau1}%
%TCIMACRO{\TeXButton{E}{\end{table}}}%
%BeginExpansion
\end{table}%
%EndExpansion
\end{center}

At each surface of the WSM ($z=0$ and $z=d$), the electric and magnetic
field must satisfy the Maxwell boundary conditions 
\begin{eqnarray}
\left( \mathbf{D}_{i}-\mathbf{D}_{f}\right) \cdot \widehat{\mathbf{n}} &=&0,
\\
\mathbf{E}_{i,\Vert }-\mathbf{E}_{f,\Vert } &=&0, \\
\left( \mathbf{B}_{i}-\mathbf{B}_{f}\right) \cdot \widehat{\mathbf{n}} &=&0,
\\
\mathbf{B}_{i,\Vert }-\mathbf{B}_{f,\Vert } &=&0,
\end{eqnarray}%
where the unit vector $\widehat{\mathbf{n}}$ points from medium $f$ to
medium $i.$ In the two dielectric media, $\mathbf{D}_{1}=n_{1}^{2}\mathbf{E}%
_{1}$ and $\mathbf{D}_{3}=n_{3}^{2}\mathbf{E}_{3}$ while in the WSM $D_{i}=%
\widetilde{\varepsilon }_{ij}E_{j}.$ The boundary conditions lead to the
system of equations

\begin{eqnarray}
\alpha _{i}+r\alpha _{r} &=&\sum_{n=1}^{n=4}t_{n}\alpha _{n};\;\alpha
_{i}-r\alpha _{r}=\sum_{n=1}^{n=4}\frac{q_{n}}{q_{i}}t_{n}\alpha _{n}, \\
\beta _{i}+r\beta _{r} &=&\sum_{n=1}^{n=4}t_{n}\beta _{n};\;\beta
_{i}-r\beta _{r}=\sum_{n=1}^{n=4}\frac{q_{n}}{q_{i}}t_{n}\beta _{n}, \\
t\alpha _{t}e^{iq_{t}d} &=&\sum_{n=1}^{n=4}t_{n}\alpha
_{n}e^{iq_{n}d}=\sum_{n=1}^{n=4}\frac{q_{n}}{q_{t}}t_{n}\alpha
_{n}e^{iq_{n}d}, \\
t\beta _{t}e^{iq_{t}d} &=&\sum_{n=1}^{n=4}t_{n}\beta
_{n}e^{iq_{n}d}=\sum_{n=1}^{n=4}\frac{q_{n}}{q_{t}}t_{n}\beta
_{n}e^{iq_{n}d}.
\end{eqnarray}%
In matrix form, and using Eqs. (\ref{q1})-(\ref{q2}) and Eq. (\ref{vec1}),
the system of equations is given by

\begin{widetext} 
\begin{equation}
\frac{1}{\sqrt{2}}\left( 
\begin{array}{cccccccc}
\sqrt{2} & 0 & -i & -i & i & i & 0 & 0 \\ 
0 & \sqrt{2} & -1 & -1 & -1 & -1 & 0 & 0 \\ 
0 & \sqrt{2} & \frac{n_{+}}{n_{1}} & -\frac{n_{+}}{n_{1}} & \frac{n_{-}}{%
n_{1}} & -\frac{n_{-}}{n_{1}} & 0 & 0 \\ 
-\sqrt{2} & 0 & -\frac{in_{+}}{n_{1}} & \frac{in_{+}}{n_{1}} & \frac{in_{-}}{%
n_{1}} & -\frac{in_{-}}{n_{1}} & 0 & 0 \\ 
0 & 0 & ie^{in_{+}\xi } & ie^{-in_{+}\xi } & -ie^{in_{-}\xi } & 
-ie^{-in_{-}\xi } & -\sqrt{2}e^{in_{3}\xi } & 0 \\ 
0 & 0 & e^{in_{+}\xi } & e^{-in_{+}\xi } & e^{in_{-}\xi } & e^{-in_{-}\xi }
& 0 & -\sqrt{2}e^{in_{3}\xi } \\ 
0 & 0 & \frac{n_{+}e^{in_{+}\xi }}{n_{3}} & -\frac{n_{+}e^{-in_{+}\xi }}{%
n_{3}} & \frac{n_{-}e^{in_{-}\xi }}{n_{3}} & -\frac{n_{-}e^{-in_{-}\xi }}{%
n_{3}} & 0 & -\sqrt{2}e^{in_{3}\xi } \\ 
0 & 0 & \frac{in_{+}e^{in_{+}\xi }}{n_{3}} & -\frac{in_{+}e^{-in_{+}\xi }}{%
n_{3}} & -\frac{in_{-}e^{in_{-}\xi }}{n_{3}} & \frac{in_{-}e^{-in_{-}\xi }}{%
n_{3}} & -\sqrt{2}e^{in_{3}\xi } & 0%
\end{array}%
\right) \left( 
\begin{array}{c}
r\alpha _{r} \\ 
r\beta _{r} \\ 
t_{1} \\ 
t_{2} \\ 
t_{3} \\ 
t_{4} \\ 
t\alpha _{t} \\ 
t\beta _{t}%
\end{array}%
\right) =\left( 
\begin{array}{c}
-\alpha _{i} \\ 
-\beta _{i} \\ 
\beta _{i} \\ 
-\alpha _{i} \\ 
0 \\ 
0 \\ 
0 \\ 
0%
\end{array}%
\right) ,  \label{system}
\end{equation}

\end{widetext}

where we have defined the parameter%
\begin{equation}
\xi =q_{0}d=\frac{\omega d}{c}.  \label{xsi}
\end{equation}

Inverting the matrix in Eq. (\ref{system}) gives the eight unknown factors
in terms of the polarization of the incident wave and the optical properties
of the three media. We remark that our procedure is not the standard way of
obtaining these factors\cite{Wu2023}. It would become cumbersome with many
interfaces. But, with only two interfaces, the $8\times 8$ matrix is easily
invertible by Mathematica or any other symbolic software.

The time-averaged energy current is given by the Poynting vector $%
\left\langle \mathbf{S}\right\rangle =\frac{1}{2\mu _{0}}\func{Re}\left[ 
\mathbf{E}\times \mathbf{B}^{\ast }\right] $ so that we have in medium $1$
and $3$%
\begin{eqnarray}
\left\langle \mathbf{S}_{1}\right\rangle \cdot \widehat{\mathbf{z}} &=&\frac{%
1}{2\mu _{0}}\frac{n_{1}}{c}E_{0}^{2}, \\
\left\langle \mathbf{S}_{3}\right\rangle \cdot \widehat{\mathbf{z}} &=&\frac{%
1}{2\mu _{0}}E_{0}^{2}\frac{n_{3}}{c}\left( \left\vert t\alpha
_{t}\right\vert ^{2}+\left\vert t\beta _{t}\right\vert ^{2}\right) .
\end{eqnarray}%
The transmission and reflection coefficients are defined by%
\begin{equation}
T=\frac{\left\langle \mathbf{S}_{3}\right\rangle _{\text{transmitted}}\cdot 
\widehat{\mathbf{z}}}{\left\langle \mathbf{S}_{1}\right\rangle _{\text{%
incident}}\cdot \widehat{\mathbf{z}}}=\frac{n_{3}}{n_{1}}\left( \left\vert
t\alpha _{t}\right\vert ^{2}+\left\vert t\beta _{t}\right\vert ^{2}\right) ,
\end{equation}%
and%
\begin{equation}
R=\frac{\left\langle \mathbf{S}_{1}\right\rangle _{\text{reflected}}\cdot 
\widehat{\mathbf{z}}}{\left\langle \mathbf{S}_{1}\right\rangle _{\text{%
incident}}\cdot \widehat{\mathbf{z}}}=\left\vert r\alpha _{t}\right\vert
^{2}+\left\vert r\beta _{t}\right\vert ^{2}
\end{equation}%
respectively.

\subsection{RCP wave}

For an incident RCP wave, the polarization vector is $\mathbf{e}_{i}=\mathbf{%
e}_{+}=\frac{1}{\sqrt{2}}\left( i,1,0\right) $ so that $\alpha _{i}=1$ and $%
\beta _{i}=1$. The polarization vector turns clockwise when looking in the
direction of propagation of the wave. (It satisfies the right-hand rule with
the thumb pointing in the direction of the axion vector $\mathbf{b}$ and the
fingers in the direction of rotation of the polarization vector\cite%
{Berman2021}.) Solving the system of equations, we get the factors%
\begin{eqnarray}
t\alpha _{t} &=&-\sqrt{2}A_{+},  \label{p1} \\
t\beta _{t} &=&\sqrt{2}iA_{+},  \label{p2}
\end{eqnarray}%
where we have defined%
\begin{equation}
A_{+}=\frac{n_{1}n_{+}e^{-in_{3}\xi }}{\left( n_{1}n_{3}+n_{+}^{2}\right)
\sin \left( n_{+}\xi \right) +i\left( n_{1}+n_{3}\right) n_{+}\cos \left(
n_{+}\xi \right) }.  \label{aplus}
\end{equation}%
Moreover, inside the WSM, we have for the factors $t_{n}$%
\begin{eqnarray}
t_{1} &=&\frac{in_{1}\left( n_{+}+n_{3}\right) e^{-in_{+}\xi }}{\left(
n_{1}n_{3}+n_{+}^{2}\right) \sin \left( n_{+}\xi \right) +n_{+}i\left(
n_{1}+n_{3}\right) \cos \left( n_{+}\xi \right) },  \label{t1} \\
t_{2} &=&\frac{in_{1}\left( n_{+}-n_{3}\right) e^{in_{+}\xi }}{\left(
n_{1}n_{3}+n_{+}^{2}\right) \sin \left( n_{+}\xi \right) +n_{+}i\left(
n_{1}+n_{3}\right) \cos \left( n_{+}\xi \right) },  \label{t2} \\
t_{3} &=&0, \\
t_{4} &=&0.
\end{eqnarray}%
The transmitted wave is also an RCP wave and only the F-RCP which is
reflected as B-LCP\ are present in the WSM.

The reflection factors are given by%
\begin{eqnarray}
r\alpha _{r} &=&\frac{i}{\sqrt{2}}C_{+}, \\
r\beta _{r} &=&\frac{1}{\sqrt{2}}C_{+},
\end{eqnarray}%
where we have defined%
\begin{equation}
C_{+}=\frac{\left( n_{1}n_{3}-n_{+}^{2}\right) \sin n_{+}\xi +n_{+}i\left(
n_{1}-n_{3}\right) \cos n_{+}\xi }{\left( n_{1}n_{3}+n_{+}^{2}\right) \sin
n_{+}\xi +n_{+}i\left( n_{1}+n_{3}\right) \cos n_{+}\xi }.
\end{equation}%
Upon reflection at the interface between regions $1$ and $2$, the RCP\ wave
becomes a LCP wave since the polarization vector does not change but the
direction of propagation is reversed.

The transmission and reflection coefficients for the RCP\ wave are given by%
\begin{eqnarray}
T_{+} &=&T_{RCP}=4\frac{n_{3}}{n_{1}}\left\vert A_{+}\right\vert ^{2}, \\
R_{+} &=&R_{RCP}=\left\vert C_{+}\right\vert ^{2}.
\end{eqnarray}

\subsection{LCP wave}

For an incident LCP wave, the polarization vector is $\mathbf{e}_{i}=\mathbf{%
e}_{-}=\frac{1}{\sqrt{2}}\left( -i,1,0\right) $ so that $\alpha _{i}=-i$ and 
$\beta _{i}=1.$ Solving the system of equations, we get%
\begin{eqnarray}
t\alpha _{t} &=&\sqrt{2}A_{-},  \label{p3} \\
t\beta _{t} &=&\sqrt{2}iA_{-},  \label{p4}
\end{eqnarray}%
where we have defined 
\begin{equation}
A_{-}=\frac{n_{1}n_{-}e^{-in_{3}\xi }}{\left( n_{1}n_{3}+n_{-}^{2}\right)
\sin \left( n_{-}\xi \right) +i\left( n_{1}+n_{3}\right) n_{-}\cos \left(
n_{-}\xi \right) }.  \label{amoins}
\end{equation}%
Inside the WSM, the factors $t_{n}$ are given by 
\begin{eqnarray}
t_{1} &=&0, \\
t_{2} &=&0, \\
t_{3} &=&\frac{in_{1}\left( n_{-}+n_{3}\right) e^{-in_{-}\xi }}{\left(
n_{1}n_{3}+n_{-}^{2}\right) \sin \left( n_{-}\xi \right) +n_{-}i\left(
n_{1}+n_{3}\right) \cos \left( n_{-}\xi \right) },  \label{t3} \\
t_{4} &=&\frac{in_{1}\left( n_{-}-n_{3}\right) e^{in_{-}\xi }}{\left(
n_{1}n_{3}+n_{-}^{2}\right) \sin \left( n_{-}\xi \right) +n_{-}i\left(
n_{1}+n_{3}\right) \cos \left( n_{-}\xi \right) }.  \label{t4}
\end{eqnarray}%
The transmitted wave is also a LCP wave. Only the F-LCP and the reflected
B-RCP\ waves are present in the WSM.

For the reflection factors, we have%
\begin{eqnarray}
r\alpha _{r} &=&\frac{-i}{\sqrt{2}}C_{-}, \\
r\beta _{r} &=&\frac{1}{\sqrt{2}}C_{-},
\end{eqnarray}%
where we have defined%
\begin{equation}
C_{-}=\frac{\left( n_{1}n_{3}-n_{-}^{2}\right) \sin n_{-}\xi +n_{-}i\left(
n_{1}-n_{3}\right) \cos n_{-}\xi }{\left( n_{1}n_{3}+n_{-}^{2}\right) \sin
n_{-}\xi +n_{-}i\left( n_{1}+n_{3}\right) \cos n_{-}\xi }.
\end{equation}

The transmission and reflection coefficients for the LCP\ wave are then 
\begin{eqnarray}
T_{-} &=&T_{LCP}=4\frac{n_{3}}{n_{1}}\left\vert A_{-}\right\vert ^{2}, \\
R_{-} &=&R_{LCP}=\left\vert C_{-}\right\vert ^{2}.
\end{eqnarray}

\subsection{LP wave}

An incident linear polarization vector can be decomposed on the RCP-LCP
basis. For $\mathbf{e}_{i}=\widehat{\mathbf{x}},$ we have $\alpha _{i}=1$
and $\beta _{i}=0$ and $\mathbf{e}_{i}=\frac{-i}{\sqrt{2}}\left( \mathbf{e}%
_{+}-\mathbf{e}_{-}\right) .$ The transmission factors are given by%
\begin{eqnarray}
t\alpha _{t} &=&i\left( A_{+}+A_{-}\right) ,  \label{p5} \\
t\beta _{t} &=&A_{+}-A_{-},  \label{p6}
\end{eqnarray}%
while we have for the reflection factors%
\begin{eqnarray}
r\alpha _{r} &=&\frac{1}{2}\left( C_{+}+C_{-}\right) , \\
r\beta _{r} &=&\frac{i}{2}\left( C_{-}-C_{+}\right) .
\end{eqnarray}%
The four factors $t_{n}$ are nonzero in this case and given by $\frac{-i}{%
\sqrt{2}}t_{1},\frac{-i}{\sqrt{2}}t_{2},\frac{i}{\sqrt{2}}t_{3},$ and $\frac{%
i}{\sqrt{2}}t_{4}.$

Because of the rotational symmetry of the WSM around the axis of propagation
(the axion term $\mathbf{b}$ being along this axis), we get for any
orientation of $\mathbf{e}_{i}$ in the $xy-$plane that the transmission and
reflection coefficients are given by%
\begin{equation}
T_{LP}=2\frac{n_{3}}{n_{1}}\left( \left\vert A_{+}\right\vert
^{2}+\left\vert A_{-}\right\vert ^{2}\right) 
\end{equation}%
and 
\begin{equation}
R_{LP}=\frac{1}{2}\left( \left\vert C_{+}\right\vert ^{2}+\left\vert
C_{-}\right\vert ^{2}\right) .
\end{equation}

\subsection{Transmission resonances}

Depending on the thickness of the WSM and the frequency of the incident
electromagnetic wave, an oscillatory pattern in the transmission
coefficients $T_{\pm }\left( \omega \right) $ can occur. Transmission
resonances in the frequency range where an RCP\ wave propagates with almost
no dissipation (\textit{i.e.} $n_{+}$ is real and positive in the clean
limit) occur when $\cos ^{2}\left( n_{+}\xi \right) =1$ or $\sin ^{2}\left(
n_{+}\xi \right) =1.$ In the first case, we have at resonance%
\begin{equation}
T_{+,1}=4\frac{n_{1}n_{3}}{\left( n_{1}+n_{3}\right) ^{2}},\newline
\end{equation}%
while in the second case 
\begin{equation}
T_{+,2}=4\frac{n_{1}n_{3}n_{+}^{2}}{\left( n_{1}n_{3}+n_{+}^{2}\right) ^{2}}.%
\newline
\end{equation}%
The maxima are given by the condition $\sin ^{2}\left( n_{+}\xi \right) =1$
if $n_{+}$ is between $n_{1}$ and $n_{3}$ in value since then $%
T_{+,2}>T_{+,1}.$ They are given by $\cos ^{2}\left( n_{+}\xi \right) =1$
otherwise. When $\cos ^{2}\left( n_{+}\xi \right) =1,$ we have inside the
WSM, for the two waves%
\begin{equation}
t_{2}=t_{1}\frac{\left( n_{+}-n_{3}\right) }{\left( n_{+}+n_{3}\right) },
\end{equation}%
while, when $\sin ^{2}\left( n_{+}\xi \right) =1,$ there is a phase shift of 
$\pi $ \textit{i.e.} 
\begin{equation}
t_{2}=-t_{1}\frac{\left( n_{+}-n_{3}\right) }{\left( n_{+}+n_{3}\right) }.
\end{equation}

At precisely $n_{1}=n_{+}$ or $n_{3}=n_{+},$ we have $T_{+,1}=T_{+,2}$ and
the oscillatory pattern in the transmission $T_{+}$ disappears. The same
remark applies to $T_{-}$ with $n_{+}$ replaced by $n_{-}.$ When $%
n_{1}=n_{3} $ (case 2 above), $T_{\pm }=1$ and $R_{\pm }=0.$ There is no
oscillation and no reflection in this case and the transmission is maximal.

The requirement that $\cos ^{2}\left( n_{\pm }\xi \right) =1$ is equivalent
to the quantization condition

\begin{equation}
d=m\frac{\lambda _{\pm }}{2},  \label{quantized}
\end{equation}%
and for $\sin ^{2}\left( n_{\pm }\xi \right) =1$ to%
\begin{equation}
d=\left( 2m+1\right) \frac{\lambda _{\pm }}{4},
\end{equation}%
where $m=1,2,3,...$ and $\lambda _{+}\left( \lambda _{-}\right) $ is the
wavelength of the RCP(LCP)\ wave \textit{inside} the WSM. The WSM\ acts as a
Fabry-P\'{e}rot interferometer in this situation.

\subsection{Faraday and Kerr rotation and ellipticity angles}

To compute the Faraday rotation and ellipticity angles for an incident
linearly polarized wave with $\mathbf{e}_{i}=\widehat{\mathbf{x}},$ we use
Eqs. (\ref{p5})-(\ref{p6}) to define the function $\eta _{F}$ by

\begin{equation}
\eta _{F}=\frac{t\beta _{t}}{t\alpha _{t}}=-i\frac{A_{+}-A_{-}}{A_{+}+A_{-}}.
\label{eta2}
\end{equation}%
In the general case where the polarization of the transmitted wave is
elliptical, the angle $\theta _{F}$ is defined as the angle that the major
axis of the polarization ellipse makes with the direction of the incident
(linear) polarization \textit{i.e.} the $x$ axis if $\mathbf{e}_{i}=\widehat{%
\mathbf{x}}.$ We use the following definition\cite{BornWolf} for the Faraday
rotation angle%
\begin{equation}
\tan 2\theta _{F}=\frac{2\func{Re}\left[ \eta _{F}\right] }{1-\left\vert
\eta _{F}\right\vert ^{2}},  \label{faradayangle}
\end{equation}%
where $\theta _{F}\in \left[ -\pi /2,\pi /2\right] .$The major and minor
axis of the ellipse have length $a$ and $g$ respectively$.$ The ellipticity
angle $\psi _{F}$ is defined\cite{BornWolf} as%
\begin{equation}
\tan \psi _{F}=\pm \frac{g}{a},  \label{ellipse}
\end{equation}%
where the $\pm $ signs indicate the direction of rotation of the electric
field vector along the ellipse and $g/a\in \left[ 0,1\right] $. Thus, a
change in the sign of the ellipticity corresponds to a change in the
direction of the rotation of the polarization vector on the ellipse. The
ellipticity angle is given by the equation\cite{BornWolf}%
\begin{equation}
\sin 2\psi _{F}=\frac{2\func{Im}\left[ \eta _{F}\right] }{1+\left\vert \eta
_{F}\right\vert ^{2}}.  \label{ellipseangle}
\end{equation}%
A linear polarization corresponds to $\psi _{F}=0$ while a circular
polarization has $\psi _{F}=-\pi /4$ for RCP\ and $\psi _{F}=\pi /4$ for LCP
(assuming forward propagation).

To compute the Kerr rotation and ellipticity angles of the reflected wave,
we assume again that the incident wave is linearly polarized with $\mathbf{e}%
_{i}=\widehat{\mathbf{x}}.$ The corresponding $\eta _{K}$ function is now 
\begin{equation}
\eta _{K}=\frac{r\beta _{r}}{r\alpha _{r}}=-i\frac{C_{+}-C_{-}}{C_{+}+C_{-}},
\end{equation}%
and thus we have%
\begin{eqnarray}
\tan 2\theta _{K} &=&\frac{2\func{Re}\left[ \eta _{K}\right] }{1-\left\vert
\eta _{K}\right\vert ^{2}},  \label{kerrangle} \\
\sin 2\psi _{K} &=&\frac{2\func{Im}\left[ \eta _{K}\right] }{1+\left\vert
\eta _{K}\right\vert ^{2}}.
\end{eqnarray}%
Because the direction of propagation is reversed, $\psi _{F}=\pi /4$ for
RCP\ and $\psi _{F}=-\pi /4$ for LCP. For any other orientation $\theta _{i}$
of the polarization vector with respect to the $x$ axis, one would need to
subtract $\theta _{i}$ from $\theta _{K}$ to obtain the effective rotation
angle $\theta _{K}-\theta _{i}$. By symmetry, this angle must be independent
of $\theta _{i}$ in the Faraday configuration.

\subsection{Thin film limit}

In the thin film limit where the wavelength of the incident light $\lambda
>>d,$ we have $\xi =$ $2.\,\allowbreak 1\times 10^{-3}<<1$ for $d=100$ nm
and $f=2\pi /\omega =10^{12}$ Hz. Thus, $\sin \left( n_{\pm }\xi \right)
\approx n_{\pm }\xi $ and $\cos \left( n_{\pm }\xi \right) \approx 1$ in $%
A_{\pm }$ and $C_{\pm }$. If we consider, in addition, that $n_{1}=n_{3},$
we have the simple result 
\begin{equation}
A_{\pm }\approx \frac{n_{1}}{\left( n_{1}^{2}+\varepsilon _{\pm }\right) \xi
+2in_{1}}.
\end{equation}%
The function $\eta _{F}<<1$ and so the Faraday and ellipticity angles are
given by 
\begin{eqnarray}
\theta _{F} &\approx &\func{Re}\left[ \eta _{F}\right] \approx \func{Re}%
\left[ \frac{n\kappa bd}{2n_{1}-i\xi \left( \varepsilon -\varepsilon
_{b}\right) }\right] ,  \label{anglecm} \\
\psi _{F} &\approx &\func{Im}\left[ \eta _{F}\right] \approx \func{Im}\left[ 
\frac{n\kappa bd}{2n_{1}-i\xi \left( \varepsilon -\varepsilon _{b}\right) }%
\right]  \notag
\end{eqnarray}%
or, to first order in $d$ by 
\begin{eqnarray}
\theta _{F} &\approx &\frac{n\kappa b\allowbreak d}{2n_{1}},
\label{tetafirst} \\
\psi _{F} &\approx &0.
\end{eqnarray}%
In this limit, the transmitted wave is linearly polarized and the Faraday
rotation is directly proportional to the WSM width, the Chern number\cite%
{Guo2023} and the axion term $b.$ We can define a rotating power by%
\begin{equation}
R_{0}=\frac{\partial \theta _{F}}{\partial d}=\frac{n\kappa b\allowbreak }{%
2n_{1}}=0.2\,\allowbreak 3\frac{nb}{n_{1}}\text{ mrad/nm,}
\end{equation}%
with $b$ in units of $10^{8}$ m$^{-1}.$

For the Kerr angle and with $n_{1}=n_{3},$ we find in the thin film limit
that%
\begin{equation}
C_{\pm }\approx \frac{\left( n_{1}^{2}-n_{\pm }^{2}\right) \xi }{\left(
n_{1}^{2}+n_{\pm }^{2}\right) \xi +2in_{1}}
\end{equation}%
and so we have%
\begin{equation}
\eta _{K}\approx \frac{2n_{1}}{\omega }\left[ \frac{nc\kappa b\allowbreak }{%
2in_{1}\left( \varepsilon -n_{1}^{2}\right) +\xi \varepsilon _{+}\varepsilon
_{-}}\right] .
\end{equation}%
However, the function $\eta _{K}$ is not small in the frequency range
considered in our calculations and so we cannot use the approximation $%
\theta _{K}\approx \func{Re}\left[ \eta _{K}\right] $ for the Kerr angle in
this limit.

\subsection{Semi-infinite WSM}

The limit of a semi-infinite WSM can be easily obtained by replacing $n_{3}$
by $n_{+}\left( \omega \right) $ in the equations for the RCP mode and by $%
n_{-}\left( \omega \right) $ for the LCP mode. We get in this way%
\begin{equation}
C_{\pm }=\frac{n_{1}-n_{\pm }}{n_{1}+n_{\pm }},
\end{equation}%
so that%
\begin{equation}
\eta _{K}=in_{1}\frac{n_{+}-n_{-}}{n_{1}^{2}-n_{+}n_{-}}.  \label{etasemi}
\end{equation}%
The reflection coefficient is given by

\begin{equation}
R_{LP}=\frac{1}{2}\left( \left\vert \frac{n_{1}-n_{+}}{n_{1}+n_{+}}%
\right\vert ^{2}+\left\vert \frac{n_{1}-n_{-}}{n_{1}+n_{-}}\right\vert
^{2}\right) .
\end{equation}%
The function $\eta _{K}$ is again not small to that the approximation $%
\theta _{K}\approx \func{Re}\left[ \eta _{K}\right] $ is not valid. The Kerr
effect for a semi-infinite WSM has been studied before. See, for example
Ref. \onlinecite{Cote2022} for Chern number $n=1$ and Ref. %
\onlinecite{Gupta2022} for mWSMs.

We remark that changing the sign of the axion term $b$ simply interchanges $%
\varepsilon _{+}$ and $\varepsilon _{-}$ and so $T_{RCP}\longleftrightarrow
T_{LCP}$ and $R_{RCP}\longleftrightarrow R_{LCP}$. The coefficients $T_{LP}$
and $R_{LP}$ do not change. The Faraday and Kerr rotation and ellipticity
angles change sign.

\section{NUMERICAL RESULTS}

\subsection{Dielectric functions}

The real (full lines) and imaginary (dashed lines)\ parts of $\varepsilon
\left( \omega \right) $ given by Eq. (\ref{epsilon}) for a WSM with two
nodes of opposite chiralities and Chern number $n=1$ is plotted in Fig. 1.
Curve 1 (black) is $\varepsilon \left( \omega \right) $ with interband
transitions only, curve 2 (blue) is $\varepsilon \left( \omega \right) $
with intraband transitions only and curve 3 (red) includes both types of
transitions. We take $\tau =100$ ps in our calculation to be in a
low-disorder limit. This parameter depends on temperature and also on the
material quality. We analyse its effect later on in this paper. Curve 4
(green) shows the effect of a finite $b_{0}$ and curve 5 (orange) shows $%
\varepsilon \left( \omega \right) $ in the absence of doping ($e_{F}=0$).
When $e_{F}\neq 0,$ the real part of $\varepsilon \left( \omega \right) $ is
positive with a logarithmic divergence at $\omega _{th}=2e_{F}/\hslash $ (%
\textit{i.e. }$f=4.83\times 10^{12}$ Hz in Fig. 1) which is the threshold
for interband absorption. When intraband transitions are absent, the
imaginary part of $\varepsilon \left( \omega \right) $ is strictly zero
below this threshold because of the Pauli blocking. With intraband
transitions considered, $\varepsilon \left( \omega \right) $ has a finite
imaginary part which is very small except at low frequency. With a finite $%
b_{0},$ the Weyl nodes are shifted in energy by $2\hslash b_{0}$ and there
are two thresholds for absorption at $\omega _{th}\pm 2b_{0}$ as shown in
Fig. 1. Intraband transitions (curve 2) make the real part of $\varepsilon
\left( \omega \right) $ negative when $\omega <\omega _{p}$ where $\omega
_{p}$ is the plasmon frequency given for $\tau \rightarrow \infty $ by%
\begin{equation}
\omega _{p}=\frac{2}{\hslash }\sqrt{\frac{n\alpha c\left( e_{F}^{2}+\hslash
^{2}b_{0}^{2}\right) }{3\pi v_{F}}},  \label{plasmon}
\end{equation}%
when only intraband transitions are considered. This frequency scales with
the square-root the of Chern number in this limit. The plasmon frequency is
however redshifted by the interband transitions as seen in curve 3. It is
then given by the solution of the transcendental equation\cite%
{Zhou2015,Cote2022}%
\begin{equation}
\omega _{p}=\sqrt{\frac{4n\alpha c}{3\pi \hslash ^{2}v_{F}}\frac{%
e_{F}^{2}+\hslash ^{2}b_{0}^{2}}{\beta \left( \omega _{p}\right) }},
\label{pplasmon}
\end{equation}%
where 
\begin{equation}
\beta \left( \omega \right) =1+\frac{n\alpha }{6v_{F}/c}\sum_{\chi }\frac{1}{%
\pi }\ln \left( \left\vert \frac{\omega ^{2}-4v_{F}^{2}k_{c}^{2}}{\omega
^{2}-4e_{F,\chi }^{2}}\right\vert \right) .
\end{equation}%
We remark in passing that the contribution of the terms that do not contain $%
e_{F,\chi }$ in the intraband part of $\varepsilon $ in Eq. (\ref{epsilon})
is negligible in the clean limit.

When interband transitions only are considered (curve 1) an electromagnetic
wave (EMW) can propagate in the WSM\ with zero attenuation if $\omega
<\omega _{th}.$ Intraband transitions introduce a region $\omega <\omega
_{p} $ where the EM wave is evanescent. For sufficiently large $\tau ,$ and
with $b=0,$ the EMW\ can propagate with almost no dissipation in the region $%
\omega \in \left[ \omega _{p},\omega _{th}\right] .$ When $e_{F}=0,$ the
EMW\ propagates at all frequencies but with some dissipation (curve 5).

In a mWSM, the conductivity tensor is multiplied by the Chern number $n$ as
indicated by Eq. (\ref{epsilon}). It follows that, when intraband
transitions only are considered, the plasmon frequency scales with $\sqrt{n}$
as indicated in Eq. (\ref{plasmon})$.$ But, when the correction $\beta
\left( \omega \right) $ is considered, the plasmon frequency becomes almost
independent of the Chern number. Indeed, as shown by curve 6 in Fig. 1, the
position of the zero of $\func{Re}\left[ \varepsilon \left( \omega \right) %
\right] $ for $n=3$ is very close to the zero of $\func{Re}\left[
\varepsilon \left( \omega \right) \right] $ for $n=1.$

\begin{figure}
\centering\includegraphics[width = \linewidth]{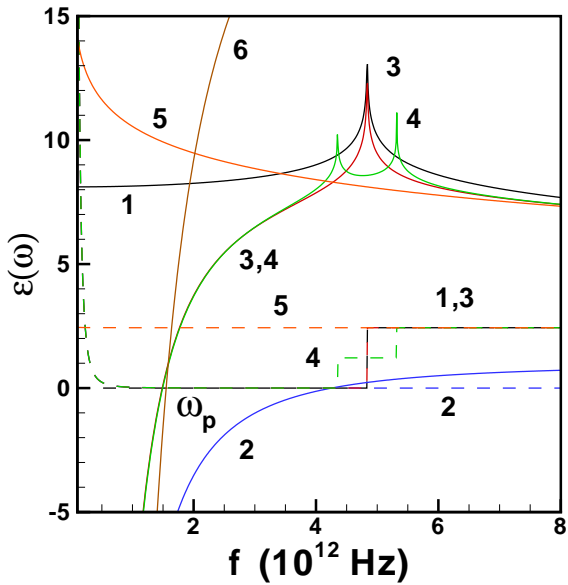} 
\caption{Fig. 1 Real (full lines) and
imaginary (dashed lines) parts of the dielectric function $\protect\varepsilon \left( \protect\omega \right) $ of a WSM\ as a function of
frequency for: (1) [black] interband transitions only, $e_{F}=10$ meV, $b_{0}=0;$ (2) [blue] intraband transitions only $e_{F}=10$ meV, $b_{0}=0;$
(3) [red] both transitions, $e_{F}=10$ meV, $b_{0}=0;$ (4) [green] both
transitions $e_{F}=10$ meV, $b_{0}=1$ meV; (5) [orange] both transitions, $e_{F}=0$ meV, $b_{0}=0$; 6 [brown] both transitions, portion of the real
part of $\protect\varepsilon \left( \protect\omega \right) $ for $e_{F}=10$
meV, $b_{0}=0$ and Chern number $n=3.$ Other parameters are $\protect\varepsilon _{b}=1,$ $v_{F}=3\times 10^{5}$ m/s, $k_{c}=50\times 10^{8}$ m$^{-1}$ and $\protect\tau =100$ ps.}
\label{fig1}
\end{figure}

When the axion term $\mathbf{b}\neq 0\mathbf{,}$ the propagation of the
electromagnetic wave depends on the dielectric functions $\varepsilon _{\pm
}\left( \omega \right) $ defined by Eqs. (\ref{q1}) and (\ref{q2}). When $%
\varepsilon _{F}\neq 0,$ $\varepsilon \left( \omega \right) $ is finite at $%
\omega =0$. In the $\omega \rightarrow 0$ limit, the asymptotic behavior of $%
\varepsilon _{\pm }\left( \omega \right) $ is governed by the sign of the
axion term i.e. $\varepsilon _{\pm }\left( \omega \rightarrow 0\right)
\rightarrow \pm \infty ~$when $b$ is positive. In this paper, however, we
concern ourselves with the behavior of the transmission and Faraday rotation
in the THz frequency domain.

The real (full lines) and imaginary (dashed lines)\ parts of the refractive
indices $n_{\pm }\left( \omega \right) $ are plotted in Fig. 2 where we have
taken $e_{F}=10$ meV, $b=1,b_{0}=0$, $\tau =100$ ps and $n=1.$ Curve 1,
however, has $b=0$ in order to indicate the position of the plasmon
frequency. Curve 2 is for $n_{-}\left( \omega \right) $ and curve 3 for $%
n_{+}\left( \omega \right) .$ The corresponding dielectric functions are
zero at $\omega _{+}$ and $\omega _{-}$ in the clean limit. These two
frequencies are defined by the solution of the transcendental equation%
\begin{equation}
\omega _{\pm }=\frac{\mp c\kappa b}{\varepsilon \left( \omega _{\pm }\right) 
}.
\end{equation}%
We find that $\omega _{+}<\omega _{p}$ and $\omega _{-}>\omega _{p}$. The
refractive indices $n_{+}\left( \omega \right) $ and $n_{-}\left( \omega
\right) $ govern the propagation of the F-RCP and F-LCP waves respectively.
When $\omega <\omega _{+},$ the F-RCP\ wave is evanescent and cannot
propagate while for $\omega \in \left[ \omega _{+},\omega _{th}\right] $ it
propagates with almost no dissipation (in the low-disorder limit). For $%
\omega >\omega _{th},$ the F-RCP\ wave propagates with some dissipation as
the refractive index is complex. The same scenario occurs for the F-LCP\
wave with $\omega _{+}$ replaced by $\omega _{-}.$ The frequency range where
the F-LCP wave propagates with almost no dissipation is smaller than that of
the F-RCP\ wave.

\begin{figure}
\centering\includegraphics[width = \linewidth]{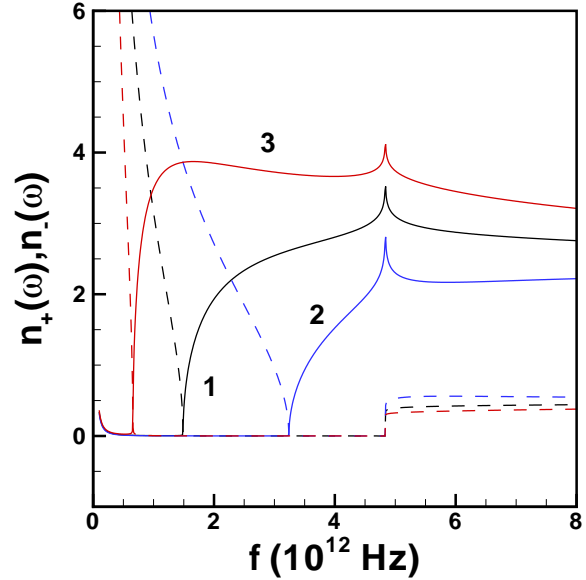} 
\caption{Real (full
lines) and imaginary part (dashed lines) of the refractive indices $n_{+}\left( \protect\omega \right) $ and $n_{+}\left( \protect\omega \right) 
$ as a function of frequencies. Curve 1 [black] $n_{\pm }\left( \protect\omega \right) ,b=0;$ 2 [blue] $n_{-}\left( \protect\omega \right) $ for $b=1\times 10^{8}$ m$^{-1};$3 [red] $n_{+}\left( \protect\omega \right) $ for 
$b=1\times 10^{8}$ m$^{-1}.$ Other parameters are $\protect\varepsilon _{b}=1,\protect\tau =100$ ps, $b_{0}=0$, $e_{F}=10$ meV, $k_{c}=50\times
10^{8}$ m$^{-1}$ and $n=1.$} \label{fig2}
\end{figure}

The lines in Fig. 3 indicate the boundaries of the propagating regions $%
\left[ \omega _{+},\omega _{th}\right] $ and $\left[ \omega _{-},\omega _{th}%
\right] $ for the F-RCP and F-LCP waves when $b=1\times 10^{8}$ m$^{-1}$ and 
$b=5\times 10^{8}$ m$^{-1}$. Below $\omega _{+}$ or $\omega _{-},$ the
corresponding wave is evanescent. Above $\omega _{th},$ both waves propagate
but with some dissipation. These frequency ranges are modified by a change
in the Fermi level and axion term $b.$ Increasing $b$ increases the
frequency range for the F-RCP wave and decreases that of the F-LCP wave. If
intraband transitions are neglected (the situation considered in Ref. %
\onlinecite{Wu2023}), then $\omega _{-}$ is slightly redshifted while $%
\omega _{+}\rightarrow 0$ so that the RCP\ wave can propagate without
dissipation in the whole frequency range $\omega \in \left[ 0,\omega _{th}%
\right] $.

\begin{figure}
\centering\includegraphics[width = \linewidth]{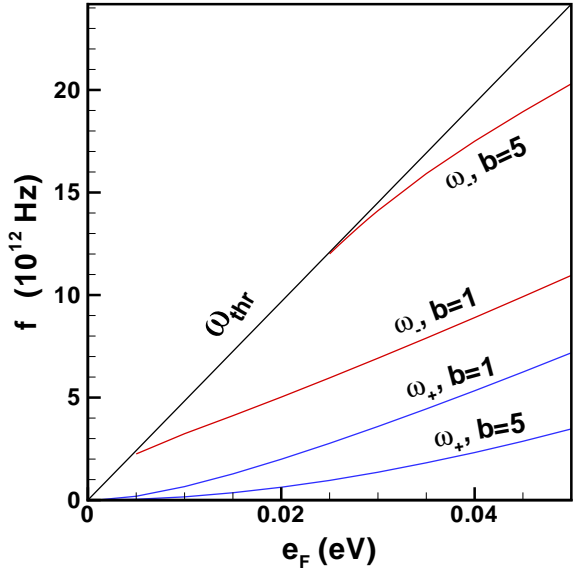} 
\caption{Behavior with Fermi level of
the boundary of the different propagating regions identified in Fig. 2 for
two values of the axion term: $b=1\times 10^{8}$ m$^{-1}$ and $b=5\times
10^{8}$ m$^{-1}.$ Other parameters are $n=1,\protect\tau =100$ ps, $\protect\varepsilon _{b}=1,$ $b_{0}=0,k_{c}=50\times 10^{8}m^{-1}.$ }
\label{fig3}
\end{figure}

\subsection{Transmission of circularly polarized waves}

We consider that the incident wave is either RCP or LCP polarized. We use
the same parameters as in Fig. 2 \textit{i.e. }$e_{F}=10$ meV, $\varepsilon
_{b}=1,b_{0}=0,b=1\times 10^{8}$ m$^{-1}$ and $\tau =100$ ps and assume a
width $d=30\mu $m for the WSM. For the two dielectrics, we take $%
n_{1}=n_{3}=1$ unless specified otherwise. The wavelength of the F-RCP\ and
F-LCP waves inside the WSM is, from Eqs. (\ref{q1}) and (\ref{q2}), given by 
\begin{eqnarray}
\lambda _{+} &=&\lambda _{F-RCP}=2\pi /\func{Re}\left[ q_{1+}\right] , \\
\lambda _{-} &=&\lambda _{F-LCP}=2\pi /\func{Re}\left[ q_{2+}\right] .
\end{eqnarray}%
Figure 4 shows the transmission coefficients $T_{RCP}$ and $T_{LCP}$ and the
ratios $d/\lambda _{+}$ and $d/\lambda _{-}$. The onsets of propagation
occurs at $d/\func{Re}\left[ \lambda _{\pm }\right] =0$ which corresponds to
the frequencies $\omega _{\pm }$. The logarithmic divergence in $d/\lambda
_{\pm }$ signals the absorption edge at $\omega _{th}.$ As discussed
previously, the RCP and LCP\ waves propagate with almost no attenuation in
the frequency range $\omega =\left[ \omega _{+},\omega _{th}\right] $ for
the RCP wave and $\omega =\left[ \omega _{-},\omega _{th}\right] $ for the
LCP wave. Above $\omega _{th},$ the waves propagate but are attenuated while
below $\omega _{\pm },$ they are evanescent. There is a small but finite
transmission even with evanescent waves if $d$ is not too big. The
transmission is finite and sizeable for each polarization when $\omega _{\pm
}<\omega <\omega _{th}.$ In these frequency ranges, there is a series of
transmission resonances which occur whenever $\cos \left( n_{\pm }\xi
\right) =1$ in $A_{\pm }$ [Eqs. (\ref{aplus}) and (\ref{amoins})]\textit{\
i.e.} whenever the quantization condition given by Eq. (\ref{quantized}) is
satisfied. If the refractive indices $n_{\pm }\left( \omega \right) $ are
real, the transmission at these resonances is maximal \textit{i.e. }$T=1.$
This maximal value is achieved when intraband transitions are neglected
since they are the only source of dissipation in our model. The transmission
drops abruptly when $\omega >\omega _{th}$.

\begin{figure}
\centering\includegraphics[width = \linewidth]{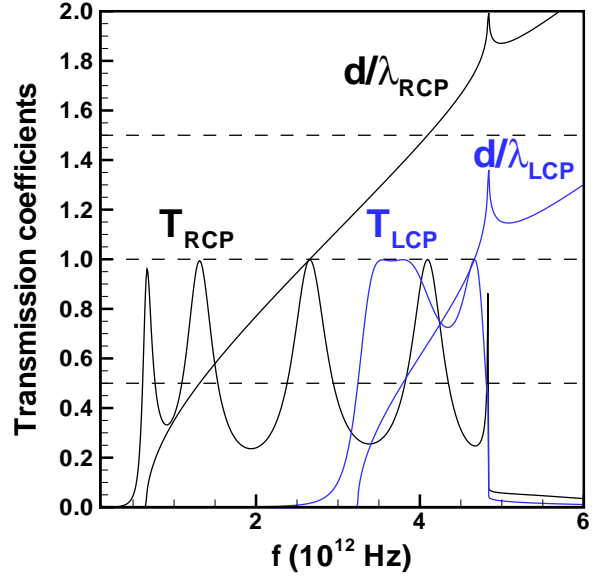} 
\caption{Transmission coefficients $T_{RCP}$ and $T_{LCP}$ and ratios $d/\protect\lambda _{RPC}$ and $d/\protect\lambda _{LCP}$ as a function of frequency for an incident RCP\ or LCP\
wave. Parameters are $n=1,$ $e_{F}=10$ meV, $b_{0}=0,\protect\tau =100$ ps, $\protect\varepsilon _{b}=1,d=30\protect\mu $m, $k_{c}=50\times 10^{8}$ m$^{-1},n_{1}=n_{3}=1,$ and $b=1\times 10^{8}$ m$^{-1}.$ The horizontal dashed
lines are set at $m/2$ where $m=1,2,3,...$ They indicate frequencies at
which the quantization condition is satisfied.} \label{fig4}
\end{figure}

Since $b$ determines in part the frequencies $\omega _{\pm }$, it follows
that the frequency range $\omega \in \left[ \omega _{\pm },\omega _{th}%
\right] $ in which the RCP and LCP\ waves propagate can be tuned by changing 
$b$ as was noticed before\cite{Wu2023}. For example, if $b$ is increased to $%
b=2.5\times 10^{8}$ m$^{-1}$ with the parameters of Fig. 4, only the RCP\
wave can be transmitted below the absorption edge. With $b=-2.5\times 10^{8}$
m$^{-1},$ the converse is true and it is the RCP\ wave that is blocked. It
is thus possible, in principle, to filter out the LCP\ wave from an incident
LP\ wave in a broad range of frequencies by changing $\mathbf{b.}$ It has
recently been shown\cite{Krizman2022} that a large enhancement of this axion
term by a factor of four can be achieved in the Weyl semimetal Cd$_{3}$As$%
_{2}$ by strain tuning.

In the frequency range $\omega \in \left[ \omega _{+},\omega _{-}\right] ,$
the RCP\ wave can propagate only forward and not backward so that a WSM\ can
also act as a broadband optical isolator\cite{Wu2023}.

As seen in Fig. 3, increasing the Fermi level $e_{F}$ for a fixed $b$ value
increases $\omega _{+}$. It is thus possible to increase the region where
both waves are blocked ($\omega <\omega _{+}$) by increasing the doping. The
Fermi level can be tuned by electric gating or by changing the temperature%
\cite{Guo2023}.

From Eq. (\ref{epsipm}), we see that increasing $\left\vert b\right\vert $
increases the refractive index $n_{+}\left( \omega \right) $ thus decreasing
the wavelength $\lambda _{+}\left( \omega \right) $ and making the
transmission coefficient $T_{RCP}\left( \omega \right) $ oscillates more
rapidly while decreasing the threshold frequency $\omega _{+}$ for the onset
of propagation. Increasing $\left\vert b\right\vert $ has the opposite
effect on $n_{-}\left( \omega \right) $.

We find numerically that increasing the Chern number increases both
refractive indices, making the oscillations more rapid and decreasing $%
\omega _{\pm }$.

\subsection{Transmission of a linearly polarized wave and Faraday rotation}

The effect of the axion term $b$ on the transmission coefficient $T_{LP}$ of
a linearly polarized (LP)\ wave is shown in Fig. 5. A linearly polarized
wave can be written as a sum of a RCP and a LCP waves of equal amplitude and
phase. When $\omega \in \left[ \omega _{+},\omega _{-}\right] ,$ only the
RCP\ portion of the LP\ wave is transmitted and, since the attenuation is
very small for $\tau =100$ ps, the maximal transmission reaches $T_{L}=0.5$
at $\omega _{+}$ and subsequently at all the transmission resonances in this
frequency range. Figure 5 also shows the Faraday ellipticity angle. It is $%
\psi _{F}=-\pi /4$ in $\omega \in \left[ \omega _{+},\omega _{-}\right] $ so
that the transmitted wave is RCP as expected. In $\omega \in \left[ \omega
_{-},\omega _{th}\right] ,$ both circular polarizations are transmitted but
with unequal amplitude. The oscillatory pattern for $T_{LP}$ is then more
complex. The transmission reaches $T_{LP}=1$ when the quantization condition
given by Eq. (\ref{quantized}) is satisfied by the two waves (with different
values of $m$ for the RCP\ and LCP components). When $T_{LP}=1$ in $\omega
\in \left[ \omega _{-},\omega _{th}\right] ,$ the transmitted wave has $\psi
_{F}=0$ and is linearly polarized. At other frequencies in this frequency
range, the polarization is elliptical. Hence, the WSM\ can act as a RCP
polarizer, in some frequency range below $\omega _{th}$ if $b>0$ or as a
LCP\ polarizer if the orientation of the WSM is reversed making $b<0$.

Above $\omega _{th},$ we can define a meaningful Faraday angle $\theta _{F}$
although the polarization is very close to being circular. The Faraday
rotation is substantial and increases with frequency while there is
simultaneously little change in the ellipticity angle $\psi _{F}$. The
transmission coefficient, however, is close to zero. A bigger transmission
coefficient is obtained by reducing the width of the WSM. This also kills
the oscillations. At $f=10^{12}$ Hz, the wavelength of light in vacuum is $%
\lambda =300\mu $m. If we choose $d=1\mu $m, then the quantization condition
for the transmission resonances is not satisfied. Figure 6 shows the
transmission coefficient $T_{LP}$, ellipticity and Faraday angles for an
incident linearly polarized wave in this limit with $b_{0}=0$ (full lines)
and $\hslash b_{0}=1$ meV (dashed lines). The transmitted wave is
elliptically polarized. The Faraday rotation angle and the transmission are
large in a broad range of frequencies. A finite $b_{0}$ creates a
logarithmic singularity in $T_{LP},\theta _{F}$ and $\psi _{F}$ at the
absorption thresholds $\omega _{th}\pm 2b_{0}$ in Fig. 6. From a measure of
the transmission coefficient, one could thus determine the energy shift
between the two nodes.

\begin{figure}
\centering\includegraphics[width = \linewidth]{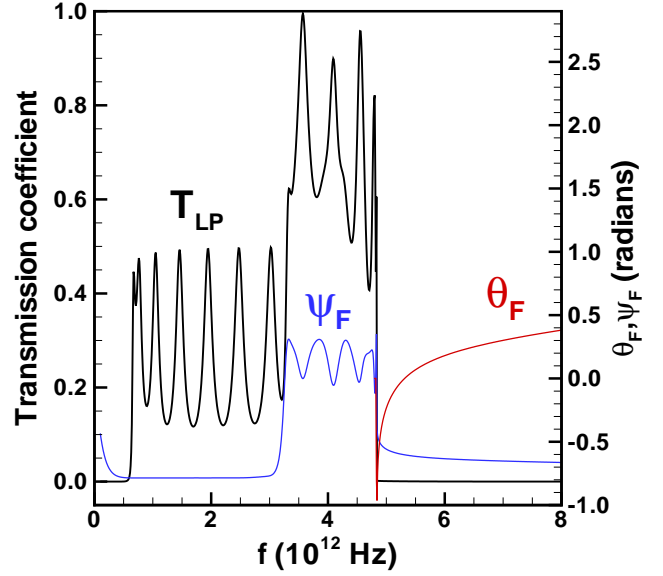} 
\caption{Transmission coefficient $T_{LP} $ and Faraday ellipticity $\protect\psi _{F}$ and rotation angle $\protect\theta _{F}$ as a function of frequency for an incident linearly
polarized wave. Parameters are $n=1,$ $e_{F}=10$ meV, $b_{0}=0,\protect\tau =100$ ps$,n_{1}=n_{3}=1,\protect\varepsilon _{b}=1,k_{c}=50\times 10^{8}$ m$^{-1}$, $b=1\times 10^{8}$ m$^{-1}$ and $d=80\protect\mu $m.}
\label{fig5}
\end{figure}

\begin{figure}
\centering\includegraphics[width = \linewidth]{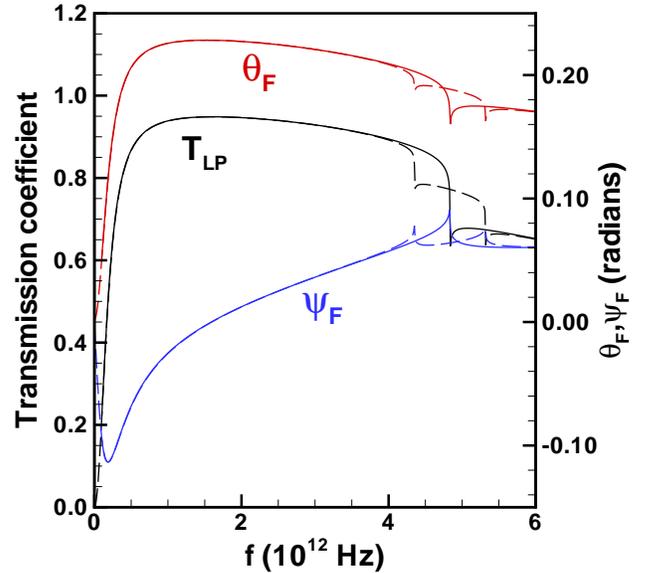} 
\caption{Transmission coefficient $T_{LP},$ ellipticity angle $\protect\psi _{F}$ and Faraday angle $\protect\theta _{F}$ for an incident linearly polarized wave as a function of
frequency. Parameters are $n=1,e_{F}=10$ meV, $\protect\tau =100$ ps, $\protect\varepsilon _{b}=1,n_{1}=n_{3}=1,b=1\times 10^{8}$ m$^{-1},k_{c}=50\times 10^{8}$ m$^{-1}$ and $d=1\protect\mu $m. Full lines: $b_{0}=0,$ dashed lines: $\hslash b_{0}=1$ meV.}
\label{fig6}
\end{figure}

In the thin-film limit, Eq. (\ref{anglecm}) shows that the Faraday and
ellipticity angles scale linearly with $b.$ Without loss of generality, we
can thus choose a frequency $f=4\times 10^{12}$ Hz from Fig. 6 and plot the
Faraday angle as a function of the width $d$ of the slab of WSM. The
resulting graph is shown in Fig. 7 for Chern number $n=1,2,3.$ The Faraday
angle increases linearly with $d$ at a rate of $R_{0}=\partial \theta
_{F}/\partial d\approx 0.21$ mrad/nm for $n=1$. This is exactly what Eqs. (%
\ref{eta2}) and (\ref{faradayangle}) give in the thin film limit $\lambda
>>d.$ Indeed, at $f=4\times 10^{12}$ Hz and for $d=100$ nm, Eq. (\ref{xsi})
gives $\xi =0.002$ and Fig. 3 shows that $n_{+}\approx 4$ and $n_{-}\approx 2
$ so that we are justified to take $\cos \left( n_{\pm }\xi \right) \approx 1
$ and $\sin \left( n_{\pm }\xi \right) \approx n_{\pm }\xi $ in Eq. (\ref%
{aplus}) and Eq. (\ref{amoins}). The Faraday angle is thus given by Eq. (\ref%
{anglecm}) (represented by the square symbols in Fig. 7). At very small $d,$
a good fit is obtained by Eq. (\ref{tetafirst}) \textit{i.e.} $\theta
_{F}\approx n\kappa b\allowbreak d/2n_{1}.$ The rotating power that we
obtain is $R_{0}\approx 0.21$ mrad/nm for $n=1$, not far from what was
recently measured\cite{Han2022} in a thin film of the magnetic WSM Co$_{2}$%
MnGa where $R_{0}\approx 3$ mrad/nm. A Faraday angle of $\theta _{F}=100$
mrad giving $R_{0}\approx 2.4$ mrad/nm was also reported\cite{Okamura2020}
for a thin film ($d=42$ nm), of another magnetic WSM: Co$_{3}$Sn$_{2}$S$_{2}$
(see Ref. \onlinecite{Kanagaraj2022} for a review of its properties). This
was considered as a large value by these authors when compared with the
conventional magneto-optical material Bi:YIG which has $R_{0}=0.77$ mrad/nm%
\cite{Okamura2020}. In yet another WSM\cite{Bandhia2020}, Co$_{2}$TiGe, the
measured Faraday angle was $\theta _{F}\approx 0.012$ rad for a width $d=72$
nm giving $R_{0}\approx 0.17$ mrad/nm. Although these three WSM\ have
different parameters, the measured Faraday angles are of the same order of
magnitude. Figure 7 shows that the range of $d$ where the linear
approximation [Eq. (\ref{tetafirst})] holds decreases with the Chern number.%
%
%\textit{\ }

Since the background dielectric constant $\varepsilon _{b}$ can be much
greater than $1$ in some WSMs, we also plot, in Fig. 7, the Faraday angle
for $\varepsilon _{b}=20$ and $n=1.$ As expected, increasing $\varepsilon
_{b}$ decreases the Faraday rotation angle at larger $d$ and reduces the
region where the linear relation is valid. This parameter has no effect in
the linear region where Eq. (\ref{tetafirst}) holds.

\begin{figure}
\centering\includegraphics[width = \linewidth]{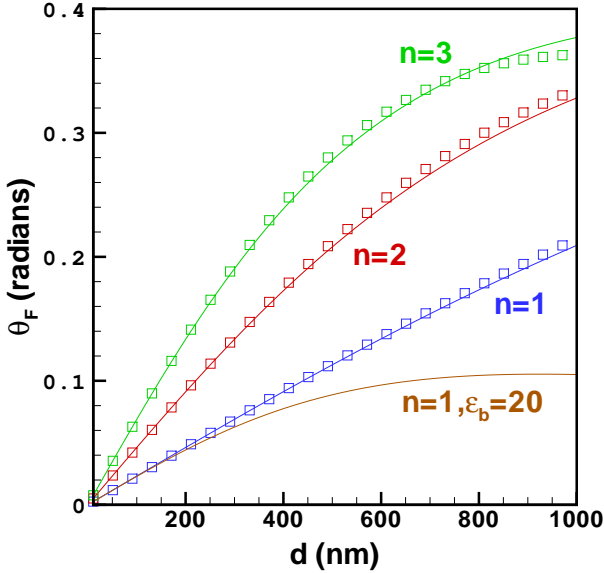} 
\caption{Faraday angles as a function of
the width $d$ of the Weyl semimetal for Chern number $n=1,2,3$ for an
incident linearly polarized wave at frequency $f=4\times 10^{12}$ Hz. The
full lines are the exact results wile the thin film limit given by Eq. (\protect\ref{anglecm}) is represented by the square symbols. Parameters are $e_{F}=10$ meV, $b_{0}=0,\protect\tau =100$ ps, $b=1\times 10^{8}$ m$^{-1},n_{1}=n_{3}=1$ and $\protect\varepsilon _{b}=1.$ The [brown] curve
labeled $n=1,\protect\varepsilon _{b}=20$ shows the effect of increasing the
background dielectric constant to $\protect\varepsilon _{b}=20.$} \label%
{fig7}
\end{figure}

Another way to control the transmitted signal is to change the dielectrics
in regions $1$ and $3.$ Figure 8 shows that the oscillations in the region $%
\omega \in \left[ \omega _{+},\omega _{-}\right] $ where only the RCP\ wave
is transmitted are completely suppressed by taking $n_{1}=n_{3}=4$ (curve 3)
which is, according to Fig. 3, the value that matches the dielectric
constant $n_{+}\left( \omega \right) $ in this frequency range. The
suppression of these oscillations is easily understood. When $%
n_{+}=n_{1}=n_{3}$ in Eq. (\ref{t2}), the coefficients of the forward and
backward propagating waves in the WSM$\ $are $t_{1}=1$ and $t_{2}=0$ so that 
$A_{+}=1/2$ and $T_{LP}=1/2.$ Since $t_{2}=0$ there is no reflected wave,
consistent with the fact that there is no discontinuity in the dielectric
constant at the two interfaces. As discussed in Sec. 4(d), the maxima of
transmission in the first four curves in Fig. 8 are given by the condition $%
\cos ^{2}\left( n_{+}\xi \right) =1.$ Curve 5 has $n_{1}=1,n_{3}=5$ and
satisfies the criteria $n_{+}\in \left[ n_{1},n_{3}\right] $ so that the
transmission maxima are given by $\sin ^{2}\left( n_{+}\xi \right) =1$ and
so shifted by half a period of oscillation.

\begin{figure}
\centering\includegraphics[width = \linewidth]{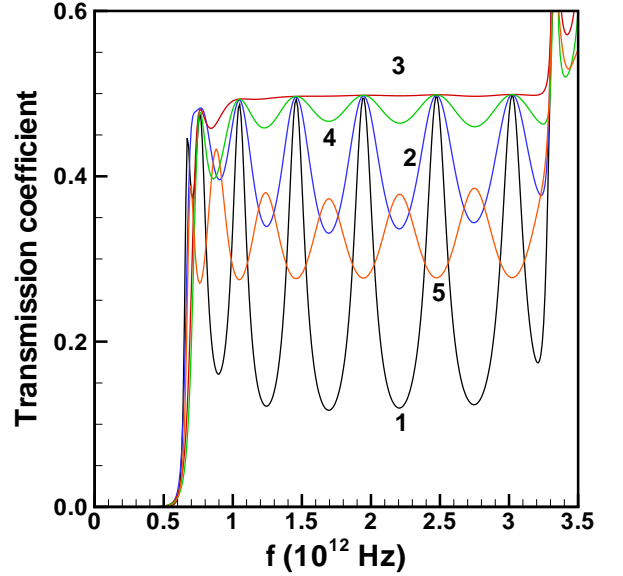} 
\caption{Effect of the dielectric
constants $n_{1}$ and $n_{3}$ on the oscillatory pattern of the transmission
coefficient $T_{LP}$ in the region where only the RCP\ wave propagates.
wave. Parameters are $e_{F}=10$ meV, $b_{0}=0,\protect\tau =100$ ps, $b=1\times 10^{8}$ m$^{-1},$ $k_{c}=50\times 10^{8},\protect\varepsilon _{b}=1,$ m$^{-1}$ and $d=80$ $\protect\mu $m. The first four curves are for $n_{1}=n_{3}$ with (1) $n_{1}=1;$ (2) $n_{1}=2;$ (3) $n_{1}=4;$ (4) $n_{1}=5$
and curve (5) has $n_{1}=1,n_{3}=5.$} \label{fig8}
\end{figure}

Figure 9 shows the threshold frequency $f_{+}=\omega _{+}/2\pi $ for the
propagation of an RCP\ wave as a function of the Fermi level for different
values of the parameters $n,\varepsilon _{b}$ and cutoff wave vector $k_{c}.$
The RCP wave propagates with almost no dissipation in the range $\Delta =%
\left[ \omega _{+},\omega _{th}\right] .$ The frequency $\omega _{th}$ is
represented by curve 1. Curve 2 shows $\omega _{+}$ and so $\Delta $ for $%
\varepsilon _{b}=1,v_{F}/c=0.001,k_{c}=50\times 10^{8}$ m$^{-1},n=1,e_{F}=10$
meV$,b_{0}=0,\tau =100$ ps$.$ With respect to curve 2, the other curves show
how $\Delta $ is modified by changing one parameter. Curve 3 shows the
effect of changing $k_{c}$ to $k_{c}=50\times 10^{8}$ m$^{-1}$ and curve 4
that of taking $n=2.$ Curve 5 has $\varepsilon _{b}=20$ and curve 6 has $%
v_{F}/c=0.003.$

\begin{figure}
\centering\includegraphics[width = \linewidth]{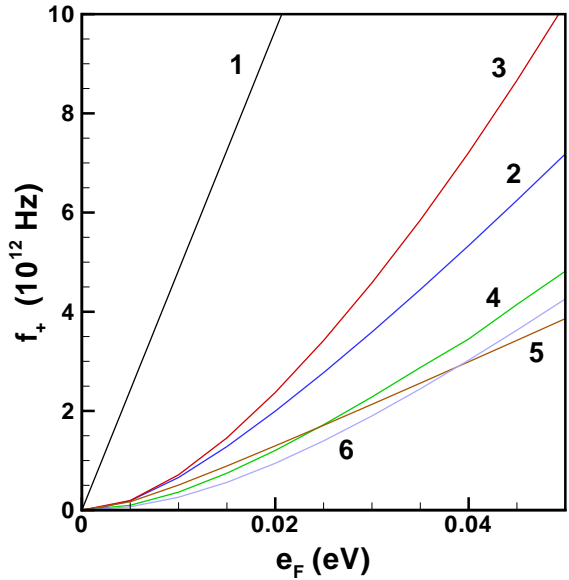} 
\caption{Effect of different parameters
on the propagation range of an RCP\ wave in the WSM. Curve 1 is the
absorption edge frequency $\protect\omega _{th}/2\protect\pi .$ The other
curves give $f_{+}=\protect\omega _{+}/2\protect\pi $ for : (2) $n=1,\protect\varepsilon _{b}=1,k_{c}=50\times 10^{8}$ m$^{-1;}$ (3) $n=1,\protect\varepsilon _{b}=1,k_{c}=5\times 10^{8}$ m$^{-1};$ (4) $n=2,\protect\varepsilon _{b}=1,k_{c}=50\times 10^{8}$ m$^{-1};$ (5) $n=1,\protect\varepsilon _{b}=20,k_{c}=50\times 10^{8}$ m$^{-1}$; (6) $n=1,\protect\varepsilon _{b}=1,k_{c}=50\times 10^{8}$ m$^{-1},v_{F}/c=0.003$. Other
parameters are $e_{F}=10$ meV, $b_{0}=0,\protect\tau =100$ ps, $b=1\times
10^{8}$ m$^{-1}$ and (for curves 1-5) $v_{F}/c=0.001$ .} \label{fig9}
\end{figure}

The scattering time $\tau $ depends on the temperature and on the quality of
the material. We have so far considered the pure limit with $\tau =100$ ps,
but $\tau $ can be substantially smaller. For example, in Co$_{2}$TiGe, it
is reported\cite{Bandhia2020} to be $\approx 0.15$ ps, in Co$_{3}$Sn$_{2}$S$%
_{2}$ it is $\approx 0.22$ ps\cite{Okamura2020} and in Co$_{2}$MnGa it is $%
\approx 0.22-0.33$ ps\cite{Han2022}. (These results are for finite
temperature.) It is thus important to see how a variation of this parameter
influences the transmission. Figure 10 shows the effect of varying the
scattering time $\tau $ on the oscillatory pattern of the transmission
coefficient for the RCP\ wave. The different curves, in order of decreasing
amplitude are for $\tau =100$ (black), $10$ (blue), $1$ (red) and $0.1$ ps
(green). The reduction in the transmission amplitude becomes really
important when $\tau $ is of the order of a picosecond or smaller. When $%
\tau <0.1,$ we find numerically that there is no crossing between $\func{Re}%
\left[ \varepsilon _{+}\left( \omega \right) \right] $ and the frequency
axis so that the region where the wave is evanescent disappears and the wave
can propagate but with some dissipation. In fact, when $\tau >0.1$ ps, $%
\func{Re}\left[ q_{1\pm }\right] <<\func{Im}\left[ q_{1\pm }\right] $ in the
frequency range where the RCP\ wave is blocked while it is the opposite if $%
\tau <0.1$ ps. This explains the increase in the transmission coefficient at
low frequency for $\tau =0.1$ ps in Fig. 10.

\begin{figure}
\centering\includegraphics[width = \linewidth]{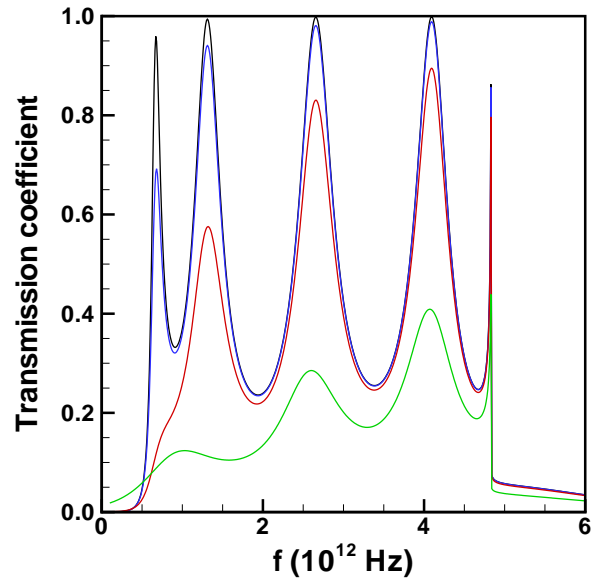} 
\caption{Effect of the scattering time $\protect\tau $ on the transmission coefficient $T_{RCP}.$ In order of
decreasing amplitude: [black] $\protect\tau =100$ ps [blue] $\protect\tau =10 $ ps; [red] $\protect\tau =1$ ps; [green] $\protect\tau =0.1$ ps
Parameters are $n=1,$ $e_{F}=10$ meV, $b_{0}=0,b=1\times 10^{8}$ m$^{-1},$ $k_{c}=50\times 10^{8},\protect\varepsilon _{b}=1,$ m$^{-1},n_{1}=n_{3}=1,$
and $d=30\protect\mu $m.} \label{fig10}
\end{figure}

\subsection{Reflection of a linearly polarized wave and Kerr rotation}

We now study the behavior of the reflection coefficient and Kerr rotation
and ellipticity angles in a WSM. Figure 11 shows the reflection coefficient
for a linearly polarized wave, $R_{LP}$ and the Kerr rotation angle $\theta
_{K}$ for a thin slab of a WSM with thickness $d=100$ nm, axion term $%
b=1\times 10^{8}$ m$^{-1}$ and for Chern number $n=1,2,3$ as indicated.
Between the two large steps in $\theta _{K}$ , the Kerr angle is close to $%
\theta _{K}=-\pi /2$ and the reflection coefficient is almost zero. We find
numerically, for the parameters used in this figure, that the first
discontinuity in $\theta _{K}$ occurs at the frequency $\omega _{+}$ and the
second at $\varepsilon _{-}\left( \omega \right) =1$ while the discontinuity
at $f\approx 5\times 10^{12}$ Hz is at the threshold frequency for optical
absorption. The calculated ellipticity (not shown) indicates that the
reflected wave in elliptically polarized in this frequency range. There is
clearly a dependency of the Kerr angle on the Chern number in this thin film
limit. Indeed, Fig. 12 shows the behavior of the Kerr angle at small
frequency for the same parameters as in Fig. 11. In the region where $\theta
_{K}$ is flat, we have clearly that $\theta _{K,n}$ for $n=2,3$ is given by $%
n\theta _{K,n=1}.$

\begin{figure}
\centering\includegraphics[width = \linewidth]{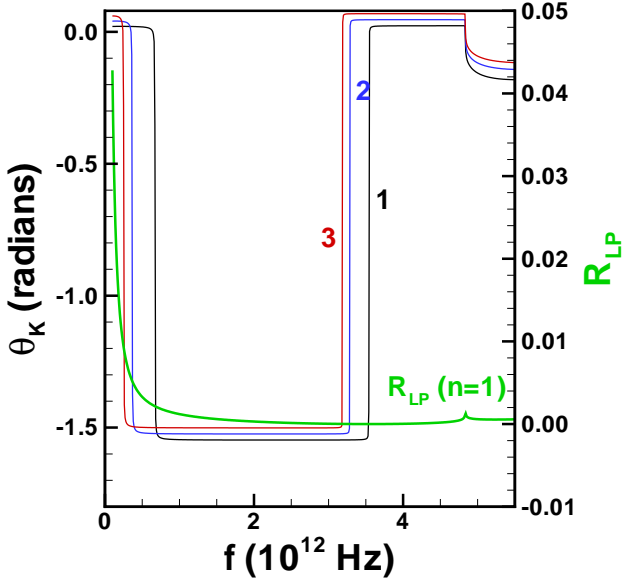} 
\caption{Reflection
coefficient $R_{LP}$ and Kerr angle $\protect\theta _{K}$ as a function of
frequency and for Chern number $n=1,2,3$ for a thin slab of a WSM with
thickness $d=100$ nm. Parameters are $e_{F}=10$ meV, $b_{0}=0,\protect\tau =100$ ps, $b=1\times 10^{8}$ m$^{-1},$ $k_{c}=50\times 10^{8},\protect\varepsilon _{b}=1,$ m$^{-1},$ $n_{1}=n_{3}=1.$ The reflection coefficient
shown is for $n=1.$} \label{fig11}
\end{figure}

In the clean limit, Fig. 6 above shows that the transmission coefficient for
the RCP\ wave in the thin-film limit is almost 1 for $\omega <\omega _{th}$.
The reflection coefficient is thus given by $R_{RCP}=1-T_{RCP}$ for $\omega
<\omega _{th}$ in the clean limit so that the oscillations in $R_{RCP}$ are
in antiphase with those of $T_{RCP}.$ The same behavior applies to the LCP
and LP\ waves. For $\omega >\omega _{th},$ however, the dissipation is
important and $T+R<1.$

\begin{figure}
\centering\includegraphics[width = \linewidth]{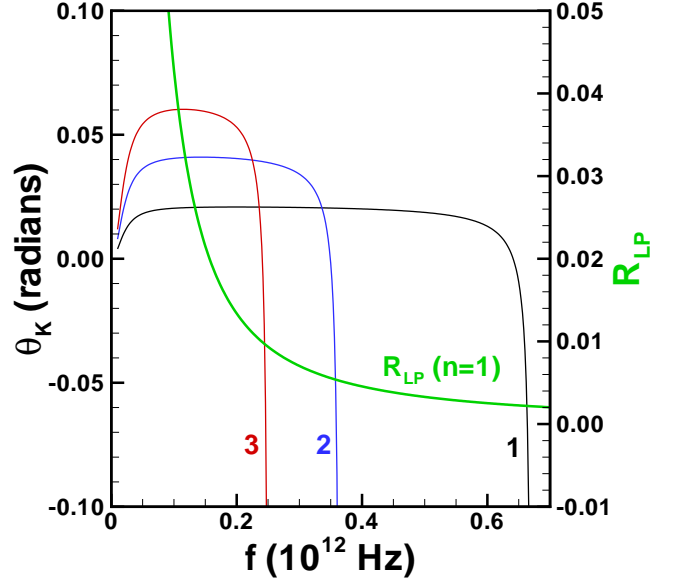} 
\caption{Reflection coefficient $R_{LP}$
and Kerr angle $\protect\theta _{K}$ at low frequency for Chern number $n=1,2,3$ for a thin slab of a WSM with thickness $d=100$ nm. The parameters
are those of Fig. 11. The reflection coefficient is shown for $n=1.$} \label%
{fig12}
\end{figure}

In Fig. 13, we show the Kerr rotation angle $\theta _{K}\left( \omega
\right) $ for Chern number $n=1,2,3$ for reflection on a semi-infinite WSM.
We use the function $\eta _{K}$ given by Eq. (\ref{etasemi}) for this graph
since the exact formula would give an infinite number of oscillations. The
Kerr rotation angle is again quite large and clearly dependent on the Chern
number.

\begin{figure}
\centering\includegraphics[width = \linewidth]{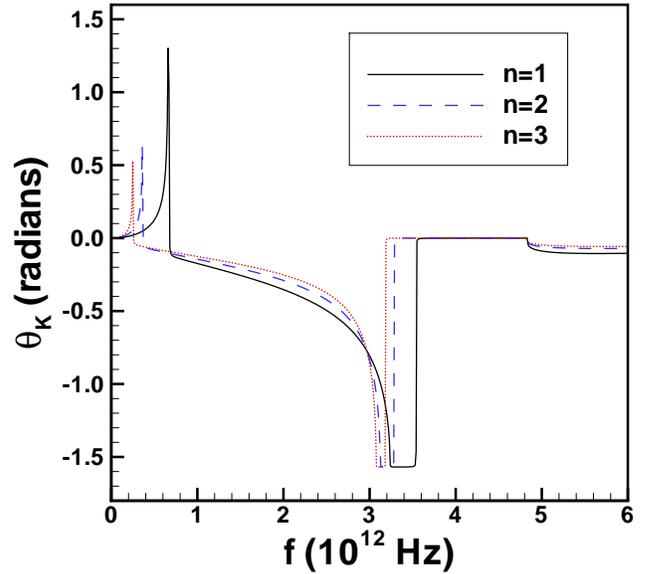} 
\caption{Kerr angle $\protect\theta _{K}\left( \protect\omega \right) $ as a function of frequency for
reflection on a semi-infinite WSM for Chern number $n=1,2,3$. Parameters are 
$e_{F}=10$ meV, $b_{0}=0,\protect\tau =100$ ps, $b=1\times 10^{8}$ m$^{-1},$ 
$k_{c}=50\times 10^{8},\protect\varepsilon _{b}=1,$ m$^{-1},$ $n_{1}=n_{3}=1. $ }
\label{fig13}
\end{figure}

Figure 14 shows the Kerr rotation and ellipticity angles and the reflection
coefficient for $n=1$ for the parameters used in Fig. 13. We find
numerically, for the parameters used in this figure, that the first
discontinuity in $\theta _{K}$ occurs at $\omega _{+}$ and the second at $%
\omega _{-}$ \textit{i.e.} at the threshold frequencies for the propagation
of the RCP and LCP waves. The discontinuity at $f\approx 5\times 10^{12}$ Hz
is at the threshold frequency for optical absorption. When $\omega
_{-}<\omega <\omega _{th},$ the dielectric functions $\varepsilon _{\pm }$
are real and positive with a negligible imaginary part so that $\func{Re}%
\left[ \eta _{K}\right] =0$. The frequency $\omega ^{\ast }$ in Fig. 14
corresponds to the condition $n_{-}=n_{1}.$ Equation (\ref{etasemi}) gives
in this case $\eta _{K}=-i$ and so $\left\vert \eta _{K}\right\vert =1.$
When $\omega _{-}<\omega <\omega ^{\ast },$ we find that $\left\vert \eta
_{K}\right\vert >1$ and Eq. (\ref{kerrangle}) gives $\theta _{K}\left(
\omega \right) =-\pi /2$. For $\omega ^{\ast }<\omega <\omega _{th,}$ we
have $\left\vert \eta _{K}\right\vert <1,$ and Eq. (\ref{kerrangle}) then
gives $\theta _{K}\left( \omega \right) =0.$The reflection coefficient is
important in this situation so that the Kerr angle is measurable. The
ellipticity changes wildly at $\omega _{\pm }$ and $\omega ^{\ast }$. We
remark that we have chosen our parameters for the Kerr angle in this section
such that $\omega _{+},\omega _{-}<\omega _{th}.$ The Kerr angle $\theta
_{K}\left( \omega \right) $ would be different if, for example, $\omega
_{+}<\omega _{th}<\omega _{-}$ as shown in Ref. \onlinecite{Cote2022}.

\begin{figure}
\centering\includegraphics[width = \linewidth]{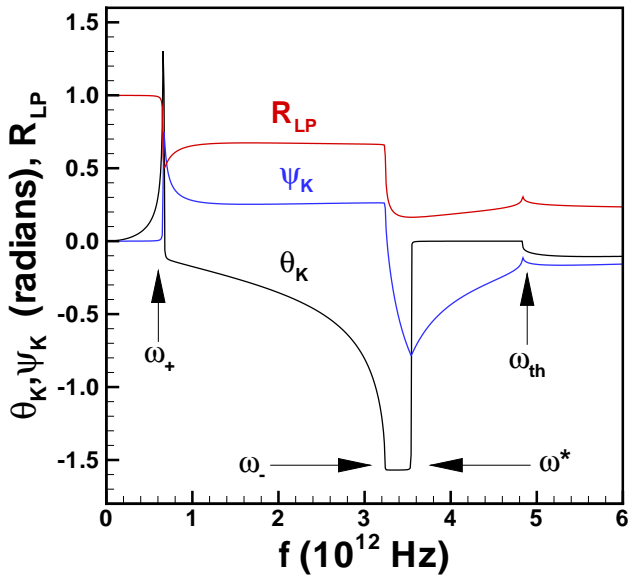} 
\caption{Kerr rotation and ellipticity
angles $\protect\theta _{K},\protect\psi _{K}$ and reflection coefficient $R_{LP}$ as a function of frequency for reflection on a semi-infinite WSM.
Parameters are $n=1,$ $e_{F}=10$ meV, $b_{0}=0,\protect\tau =100$ ps, $b=1\times 10^{8}$ m$^{-1},$ $k_{c}=50\times 10^{8},\protect\varepsilon _{b}=1,$ m$^{-1},$ $n_{1}=n_{3}=1.$ }
\label{fig14}
\end{figure}

\section{CONCLUSION}

The Maxwell equations that govern the propagation of electromagnetic waves
inside a Weyl semimetal with broken time-reversal and space inversion
symmetries are modified by the presence of the axion terms $\mathbf{b}$ and $%
b_{0}.$ When light is incident on the surface of a WSM\ with no Fermi arcs
and propagates in the direction of the vector $\mathbf{b},$ the right and
left circular polarization experience different refractive indices. Hence,
they propagate with different speeds and attenuation. The threshold
frequency for their propagation is also different. (The waves are evanescent
below the threshold frequency.) The axion term $\mathbf{b}$ (and to some
lesser extent $b_{0}$) also confers a gyrotropic nature to a WSM\ so that a
linearly polarized wave will experience a rotation of its polarization
vector upon reflection (Kerr effect) or transmission (Faraday effect)\cite%
{Kargarian,Ghosh2023,Wu2023,Yang2022,Berman2021}. Thus, at a fundamental
level, a WSM is non-reciprocal medium with birefringent and dichroic
characters even in the absence of an external magnetic field. At a practical
level, WSMs can be used to make optical devices that act as chiral filters
or polarizers or as optical isolators\cite{Guo2023} since, in this later
case, it is possible to block the transmission of the RCP or LCP waves in a
broad range of frequencies.

In this paper, we made an in-depth study of these effects by analyzing in
detail how the different parameters that characterize a WSM control these
unique optical properties. Working with a simple two-node model, but
allowing for the Chern number of the two nodes to be $n=1,2,3,$ we studied
the transmission, reflection and Kerr and Faraday effects for light incident
on a thin film, a slab or a semi-infinite WSM. In a slab, there is the
additional effect that the multiple reflections inside the WSM\ lead to
transmission resonances at some quantized frequencies. The WSM then acts as
a Fabry-P\'{e}rot interferometer. We have considered the following
parameters in our study: axion terms, Fermi level (doping), Fermi velocity,
Chern number, width of the WSM, relaxation time for intraband scattering,
size of the high-energy cutoff, background dielectric constant and
refractive indices of the dielectrics on both sides of the slab of WSM.

We limited our study to zero temperature. A finite temperature will change
the Fermi level and therefore the numerical values of the different
threshold frequencies that control the propagation of the electromagnetic
waves inside the WSM. We did not consider a tilting of the Weyl nodes. Its
effects have been studied extensively in the past years and more recently in
multi-WSMs\cite{Liu2022,Gupta2022,Carbotte2018,Yadav2023,Das2022}.

The Faraday configuration that we studied in this paper is associated with
the circularly polarized waves. In the Voigt configuration where the
propagation is perpendicular to $\mathbf{b},$ the eigenmodes are linearly
polarized. It was found in this case that a wave is transmitted if its
polarization is collinear to $\mathbf{b}$ and reflected if it is
perpendicular to $\mathbf{b}$\cite{Berman2021}. In this configuration,
however, Fermi arcs that are present on the surfaces parallel to $\mathbf{b}$
should, in principle, be considered in the analysis.

\begin{acknowledgments}
R. C\^{o}t\'{e} was supported by a grant from the Natural Sciences and
Engineering Research Council of Canada (NSERC). R. N. Duchesne and G. D.
Duchesne were supported by a scholarship from NSERC and the Fonds de
recherche du Qu\'{e}bec-Nature et technologies (FRQNT).
\end{acknowledgments}

\smallskip \appendix

\section{CONDUCTIVITY OF A WSM\ WITH CHERN NUMBER $n=1,2,3$}

In this appendix, we show that considering a WSM\ with a higher Chern number 
$n=2,3$ only multiplies the diagonal elements of the conductivity $\sigma
_{xx},\sigma _{yy}$ by the topological charge $n$ when the Hamiltonian is
that given by Eq. (\ref{hamilton}) where the linear dispersion is along the $%
z$ axis. The conductivity element $\sigma _{zz}$ gets a more complex
dependence\cite{Conductivite} on $n$ but, as it does not enter in our
calculations of the transmission coefficients and Faraday angles, we will
not discuss it here.

The current operator is obtained by making the Peierls substitution $\mathbf{%
k}\rightarrow \mathbf{k}+\frac{e}{\hslash }\mathbf{A}$ in the Hamiltonian
and then taking the derivative $j_{\alpha }=-\frac{\delta H}{\delta
A_{\alpha }},$ where $\alpha =x,y,z.$ In cylindrical coordinates $\left(
\varphi ,k_{\bot },k_{z}\right) $, we get for $j_{x}$ and $j_{y}$%
\begin{eqnarray}
j_{x} &=&-\chi nv_{F}\beta e\left( 
\begin{array}{cc}
0 & \left( k_{\bot }e^{-i\varphi }\right) ^{n-1} \\ 
\left( k_{\bot }e^{i\varphi }\right) ^{n-1} & 0%
\end{array}%
\right) , \\
j_{y} &=&-\chi nv_{F}\beta ei\left( 
\begin{array}{cc}
0 & -\left( k_{\bot }e^{-i\varphi }\right) ^{n-1} \\ 
\left( k_{\bot }e^{i\varphi }\right) ^{n-1} & 0%
\end{array}%
\right) ,
\end{eqnarray}%
where we have omitted the diamagnetic contributions. In these equations, $%
\tau $ is the node index and $n$ the Chern number. With the eigenvectors
given by Eq. (\ref{eigenv}), the matrix elements of the current operator are
for node $\chi =+1$ and band index $s=\pm 1$%
\begin{eqnarray}
\left\langle \eta _{+}\left( \mathbf{k}\right) \right\vert j^{\left(
x\right) }\left\vert \eta _{-}\left( \mathbf{k}\right) \right\rangle
&=&-nv_{F}\beta ek_{\bot }^{n-1} \\
&&\times \frac{k_{z}\cos \varphi +i\sin \varphi \sqrt{k_{z}^{2}+\beta
^{2}k_{\bot }^{2n}}}{\sqrt{k_{z}^{2}+\beta ^{2}k_{\bot }^{2n}}},  \notag \\
\left\langle \eta _{+}\left( \mathbf{k}\right) \right\vert j^{\left(
y\right) }\left\vert \eta _{-}\left( \mathbf{k}\right) \right\rangle
&=&-nv_{F}\beta ek_{\bot }^{n-1} \\
&&\times \frac{k_{z}\sin \varphi -i\cos \varphi \sqrt{k_{z}^{2}+\beta
^{2}k_{\bot }^{2n}}}{\sqrt{k_{z}^{2}+\beta ^{2}k_{\bot }^{2n}}}.  \notag
\end{eqnarray}%
For $\chi =-1,$ it suffices to make the substitution $\left\vert \eta _{\pm
}\left( \mathbf{k}\right) \right\rangle \rightarrow \left\vert \eta _{\mp
}\left( \mathbf{k}\right) \right\rangle $ and multiply the result by $-1.$%
The WSM\ being isotropic around the $z$ axis, we need, for the current
response function, the terms (written here for $\tau =1$)

\begin{widetext}

\begin{eqnarray}
\int_{0}^{2\pi }\frac{d\varphi }{2\pi }\left\langle \eta _{+}\left( \mathbf{k%
}\right) \right\vert j^{\left( x\right) }\left\vert \eta _{-}\left( \mathbf{k%
}\right) \right\rangle \left\langle \eta _{-}\left( \mathbf{k}\right)
\right\vert j^{\left( x\right) }\left\vert \eta _{+}\left( \mathbf{k}\right)
\right\rangle  &=&n^{2}v_{F}^{2}\beta ^{2}e^{2}\frac{k_{\bot }^{2n-2}}{%
k_{z}^{2}+\beta ^{2}k_{\bot }^{2n}}\left( k_{z}^{2}+\frac{1}{2}\beta
^{2}k_{\bot }^{2n}\right) ,  \label{a1} \\
\int_{0}^{2\pi }\frac{d\varphi }{2\pi }\left\langle \eta _{+}\left( \mathbf{k%
}\right) \right\vert j^{\left( y\right) }\left\vert \eta _{-}\left( \mathbf{k%
}\right) \right\rangle \left\langle \eta _{-}\left( \mathbf{k}\right)
\right\vert j^{\left( y\right) }\left\vert \eta _{+}\left( \mathbf{k}\right)
\right\rangle  &=&n^{2}v_{F}^{2}\beta ^{2}e^{2}\frac{k_{\bot }^{2n-2}}{%
k_{z}^{2}+\beta ^{2}k_{\bot }^{2n}}\left( k_{z}^{2}+\frac{1}{2}\beta
^{2}k_{\bot }^{2n}\right) ,  \label{a2} \\
\int_{0}^{2\pi }\frac{d\varphi }{2\pi }\left\langle \eta _{+}\left( \mathbf{k%
}\right) \right\vert j^{\left( x\right) }\left\vert \eta _{-}\left( \mathbf{k%
}\right) \right\rangle \left\langle \eta _{-}\left( \mathbf{k}\right)
\right\vert j^{\left( y\right) }\left\vert \eta _{+}\left( \mathbf{k}\right)
\right\rangle  &=&n^{2}v_{F}^{2}\beta ^{2}e^{2}k_{\bot }^{2n-2}\frac{ik_{z}}{%
\sqrt{k_{z}^{2}+\beta ^{2}k_{\bot }^{2n}}}.  \label{a3}
\end{eqnarray}%
The response function $\chi _{x,x}\left( \omega \right) $ for node $\chi =1$
is%
\begin{eqnarray}
\chi _{x,x}\left( \omega \right)  &=&-\frac{1}{\hslash }\int_{0}^{2\pi }%
\frac{d\varphi }{2\pi }\int_{-k_{c}}^{+k_{c}}dk_{z}\int_{0}^{\infty }\frac{%
dk_{\bot }}{\left( 2\pi \right) ^{2}}k_{\bot }\left\langle \eta _{+}\left( 
\mathbf{k}\right) \right\vert j^{\left( x\right) }\left\vert \eta _{-}\left( 
\mathbf{k}\right) \right\rangle \left\langle \eta _{-}\left( \mathbf{k}%
\right) \right\vert j^{\left( x\right) }\left\vert \eta _{+}\left( \mathbf{k}%
\right) \right\rangle  \\
&&\times \left[ \frac{\left\langle n_{-1,\mathbf{k}}\right\rangle
-\left\langle n_{1,\mathbf{k}}\right\rangle }{\omega +i\delta +2E\left( 
\mathbf{k}\right) /\hslash }+\frac{\left\langle n_{1,\mathbf{k}%
}\right\rangle -\left\langle n_{-1,\mathbf{k}}\right\rangle }{\omega
+i\delta -2E\left( \mathbf{k}\right) /\hslash }\right]   \notag \\
&=&-e^{2}v_{F}^{2}\frac{1}{\hslash }n^{2}\beta
^{2}\int_{-k_{c}}^{+k_{c}}dk_{z}\int_{0}^{\infty }\frac{dk_{\bot }}{\left(
2\pi \right) ^{2}}\frac{k_{\bot }^{2n-1}\left( k_{z}^{2}+\frac{1}{2}\beta
^{2}k_{\bot }^{2n}\right) }{k_{z}^{2}+\beta ^{2}k_{\bot }^{2n}}\Theta \left(
\hslash v_{F}\sqrt{k_{z}^{2}+\beta ^{2}k_{\bot }^{2n}}-e_{F,1}\right)  
\notag \\
&&\times \left[ \frac{1}{\omega +i\delta +2E\left( \mathbf{k}\right)
/\hslash }-\frac{1}{\omega +i\delta -2E\left( \mathbf{k}\right) /\hslash }%
\right] ,  \notag
\end{eqnarray}%
where $\left\langle n_{s,\mathbf{k}}\right\rangle $ is the occupation of
level $\mathbf{k}$ in band $s$ and the energy is 
\begin{equation}
E\left( \mathbf{k}\right) =\hslash v_{F}\sqrt{k_{z}^{2}+\beta ^{2}k_{\bot
}^{2n}}.
\end{equation}

We work at $T=0$ K and assume that the Fermi level is in band $s=+1$ so that
the occupation factors are%
\begin{eqnarray}
\left\langle n_{-1,\mathbf{k}}\right\rangle &=&1, \\
\left\langle n_{1,\mathbf{k}}\right\rangle &=&\Theta \left( e_{F,\tau
}-E\left( \mathbf{k}\right) \right) .  \notag
\end{eqnarray}%
From Eqs. (\ref{a1}) and (\ref{a2}), we have $\chi _{x,x}\left( \omega
\right) =\chi _{y,y}\left( \omega \right) $ and since $E\left( \mathbf{k}%
\right) $ is even in $k_{z},$ Eq. (\ref{a3}) gives $\chi _{x,y}\left( \omega
\right) =\chi _{y,x}\left( \omega \right) =0$ when the two nodes are
considered as two separate systems (each node is assume to be centered at $%
k_{z}=0$).

If we make the change of variables 
\begin{eqnarray}
y &=&v_{F}\beta k_{\bot }^{n}, \\
x &=&v_{F}k_{z},
\end{eqnarray}%
we get for both nodes%
\begin{eqnarray}
\chi _{x,x}\left( \omega \right)  &=&-\frac{e^{2}n}{\left( 2\pi \right)
^{2}\hslash v_{F}}\sum_{\chi
}\int_{-v_{F}k_{c}}^{+v_{F}k_{c}}dx\int_{0}^{\infty }dy\frac{y\left( x^{2}+%
\frac{1}{2}y^{2}\right) }{x^{2}+y^{2}}\Theta \left( \sqrt{x^{2}+y^{2}}%
-e_{F,\chi }/\hslash \right)  \\
&&\times \left[ \frac{1}{\omega +i\delta +2\sqrt{x^{2}+y^{2}}}-\frac{1}{%
\omega +i\delta -2\sqrt{x^{2}+y^{2}}}\right] .  \notag
\end{eqnarray}%
As we see, the integrals are independent of the Chern number $n$ which
appears only as a multiplicative factor.

If we shift the nodes by $\pm b$ along the $k_{z}$ axis but still integrate
from $-k_{c}$ to $k_{c}$ for each node, we get a nonzero result for $\sigma
_{xy}\left( \omega \right) $ which is given by\cite{Sonowal2019} (for $%
b_{0}=0$)

\begin{eqnarray}
\func{Re}\left[ \sigma _{xy}\left( \omega \right) \right] &=&\frac{ne^{2}}{%
2\pi h}b  \label{sigma1} \\
&&+\frac{ne^{2}}{16v_{F}\omega \pi h}\left(
4b^{2}v_{F}^{2}+8bv_{F}^{2}k_{c}-\omega ^{2}+4v_{F}^{2}k_{c}^{2}\right) \ln
\left( \frac{2v_{F}k_{c}+2v_{F}b+\omega }{2v_{F}k_{c}+2v_{F}b-\omega }\right)
\notag \\
&&-\frac{ne^{2}}{16\omega v_{F}\pi h}\left(
4b^{2}v_{F}^{2}-8bv_{F}^{2}k_{c}-\omega ^{2}+4v_{F}^{2}k_{c}^{2}\right) \ln
\left( \frac{2v_{F}k_{c}-2v_{F}b+\omega }{2v_{F}k_{c}-2v_{F}b-\omega }%
\right) ,  \notag
\end{eqnarray}%
or, to order $\omega ^{2}$ by 
\begin{equation}
\func{Re}\left[ \sigma _{xy}\left( \omega \right) \right] =\frac{ne^{2}b}{%
\pi h}+\frac{1}{12}\frac{ne^{2}}{\pi h}\frac{\omega ^{2}b}{%
v_{F}^{2}k_{c}^{2}-v_{F}^{2}b^{2}},  \label{sigma2}
\end{equation}%
or for the simplest approximation%
\begin{equation}
\func{Re}\left[ \sigma _{xy}\left( \omega \right) \right] =\frac{ne^{2}b}{%
\pi h},  \label{sigma3}
\end{equation}%
and $\func{Im}\left[ \sigma _{xy}\left( \omega \right) \right] =0$ for $%
\omega <2e_{F}/\hslash $ which is the frequency range where interesting
effects occur in the transmission. With the parameters $v_{F}=3\times 10^{5}$
m/s, $k_{c}=50\times 10^{8}$ m$^{-1},$ we find that the difference between
the 3 expressions for $\func{Re}\left[ \sigma _{xy}\left( \omega \right) %
\right] $ is less than $0.0001$ for $b=1\times 10^{8}$ m$^{-1}$ and $0.001$
for $b=10\times 10^{8}$ m$^{-1}$ in the frequency range $f\in \left[ 1,10%
\right] \times 10^{12}$ Hz. We can thus safely use the simplest result given
by Eq. (\ref{sigma3}) in our calculations.

\end{widetext}

\end{document}